\documentclass[aps,floats]{revtex4}
\usepackage{amsmath,amssymb}
\usepackage{graphicx,epsfig}

\begin{document}
\bibliographystyle {plain}

\def\oppropto{\mathop{\propto}} 
\def\opmin{\mathop{\min}} 
\def\opmax{\mathop{\max}} 
\def\opsimeq{\mathop{\simeq}}
\def\opoverderline{\mathop{\overline}}
\def\operarrow{\mathop{\longrightarrow}}
\def\opsim{\mathop{\sim}} 
\def\oplim{\mathop{\lim}} 

\def\fig#1#2{\includegraphics[height=#1]{#2}}
\def\figx#1#2{\includegraphics[width=#1]{#2}}


\title{ Non-equilibrium steady states : maximization of the Shannon entropy associated \\
 with the distribution of dynamical trajectories in the presence of constraints } 


 \author{ C\'ecile Monthus }
  \affiliation{ Institut de Physique Th\'{e}orique, CNRS and CEA Saclay,
 91191 Gif-sur-Yvette, France}

\begin{abstract}
Filyokov and Karpov [Inzhenerno-Fizicheskii Zhurnal 13, 624 (1967)] have proposed a theory of non-equilibrium steady states in direct analogy with the theory of equilibrium states : the principle is to maximize the Shannon entropy associated with the probability distribution of dynamical trajectories in the presence of constraints, including the macroscopic current of interest, via the method of Lagrange multipliers. This maximization leads directly to  generalized Gibbs distribution for the probability distribution of dynamical trajectories, and to some fluctuation relation of the integrated current. The simplest stochastic dynamics where these ideas can be applied are discrete-time Markov chains, defined by transition probabilities $W_{i \to j}$ between configurations $i$ and $j$ : instead of choosing the dynamical rules $W_{i \to j} $ a priori, one determines the transition probabilities and the associate stationary state that maximize the entropy of dynamical trajectories with the other physical constraints that one wishes to impose. We give a self-contained and unified presentation of this type of approach,  both for discrete-time Markov Chains and for continuous-time  Master Equations. The obtained results are in full agreement with the Bayesian approach introduced by Evans [Phys. Rev. Lett. 92, 150601 (2004)] under the name 'Non-equilibrium Counterpart to detailed balance', and with the 'invariant quantities' derived by Baule and Evans [Phys. Rev. Lett. 101, 240601 (2008)], but provide a slightly different perspective via the formulation in terms of an eigenvalue problem.

\end{abstract}

\maketitle

 \section{ Introduction} 

In contrast to the statistical physics of equilibrium, which is based on a few general principles, everybody complains that 
no global theory has emerged to describe out-of-equilibrium systems.
A natural explanation of this matter of fact is that the problem is ill-defined
for many reasons. A first reason is that
there are actually very different ways to be out-of-equilibrium :
a system can be 'out-of-equilibrium' because it has a very slow dynamics 
that never converges towards some stationary state, or because it is 
in a non-equilibrium steady state, characterized by some currents
that are imposed by the environment.
It is clear that in the first case, one really needs a non-trivial dynamical description,
whereas in the second case, one may hope that the 'steady state' property 
could help to develop some theory generalizing the equilibrium steady state.
A second reason is that
even within the field of 'non-equilibrium steady states', 
very different goals have been pursued. 
On one hand, many theorists of statistical physics consider 
that the best strategy consists in solving specific models with chosen dynamical rules,
in order to see if some general principles emerge in the end :
in this direction, the most studied models are one dimensional exclusion processes,
for which many remarkable results have been obtained over the years (see the recent short review \cite{derrida_2010} and references therein).
On the other hand, it is clear that the success of the theory of equilibrium
is based on the fact that one does not specify the details of the dynamical 
evolution but that on the contrary, one keeps only its conservation properties
(for instance the energy conservation) 
and then maximize the Shannon entropy associated with the equilibrium state.
So there has been various attempts to formulate general principles based on 
some notion of 'entropy' for non-equilibrium steady states.
Although the principles of 'minimum entropy production'
and of 'maximum entropy production' seem popular, in particular
in applications to related areas like chemistry, biology or climatology,
the proposed justifications have not been widely accepted,
either because there are at the level of 'generalized thermodynamics'
rather than statistical physics, 
or because they are limited to 'near-equilibrium' conditions,
or because they involve other additional questionable assumptions
 (see for instance the discussions in
 \cite{revueprodentropie,bruers,minentropyprod,elec} and references therein).
But we feel that instead of focusing on the notion of 'entropy production' to formulate general principles,
it is much more natural to consider  the 'Shannon entropy associated with the probability distribution
of dynamical trajectories' that we will call equivalently 'dynamical entropy'
 to have a shorter name. This notion can be introduced for any
 stochastic dynamics, or for any deterministic chaotic dynamics that becomes effectively 'stochastic',
 as follows : to characterize the dynamics during some time interval $[0,t]$,
one considers all possible dynamical trajectories $\Omega_{[0,t]}$ 
and their probabilities ${\cal P}(\Omega_{[0,t]})$ normalized to
\begin{eqnarray}
\sum_{\Omega_{[0,t]}} {\cal P}(\Omega_{[0,t]}) =1
\label{normatraj}
\end{eqnarray}
The Shannon entropy associated with this distribution ${\cal P}(\Omega_{[0,t]})$
over dynamical trajectories reads
\begin{eqnarray}
S^{dyn}(t) \equiv - \sum_{\Omega_{[0,t]}}
 {\cal P}(\Omega_{[0,t]}) \ln {\cal P}(\Omega_{[0,t]}) 
\label{straj}
\end{eqnarray}
For many stochastic process of interest, this entropy becomes
extensive in the time $t$ in the large-time limit $t \to +\infty$ :
 it is then convenient to introduced the entropy rate per unit time, called the
Kolmogorov-Sinai entropy in the context of chaotic dynamical systems
(see for instance the book \cite{book_gaspard} and references therein)
\begin{eqnarray}
h_{KS} \equiv \oplim_{t \to + \infty} \frac{S^{dyn}(t)}{t} 
\label{hKS}
\end{eqnarray}

More than forty years ago, Filyokov and Karpov
\cite{russes,russeseq,russesneareq} have proposed 
to base the theory of non-equilibrium steady states on the maximization
of the dynamical entropy of Eq. \ref{straj} in the presence
of some constraints via the method of Lagrange multipliers.
We are not aware of direct continuations of these works,
except the very recent work of Favretti \cite{favretti}.
However closely related ideas have been proposed independently by Evans 
\cite{evans} under the name 'Non-equilibrium Counterpart to detailed balance'
and have been applied to shear flow \cite{evansinvariants,evansetal,evansreview}.
Also among recent attempts to 'prove' the principle of maximum entropy production by
Dewar \cite{dewar}, one can see that the starting point is precisely the maximization
of the dynamical entropy of Eq. \ref{straj}.
We thus feel that if one wishes to formulate
a general principle based on the maximization of some notion of entropy for
non-equilibrium steady states, the dynamical entropy of Eq. \ref{straj}
is definitely the most natural and could be the basis of some consensus.
 
Let us now mention various recent works that are related to some of these notions :

- For random walks on arbitrary networks, the optimization of the
dynamical entropy of Eq. \ref{straj} leads to the 'maximal entropy random walk' 
that has been introduced by Burda, Duda, Luck and Waclaw \cite{jmluck} : 
it is different from the usual random walk that maximizes the entropy only locally
whenever there exists 
some fluctuations in the coordination numbers of nodes of the network
(but of course these two types of random walks coincide on regular networks).

- For non-equilibrium steady states,
even within the point of view where the dynamical rules are given a priori,
the dynamical entropy of Eq. \ref{straj} 
plays a major role in the 'thermodynamics of histories'
that has been developed by Lecomte, Appert-Rolland and Van Wijland \cite{vivien},
and that has been then applied to various glassy systems \cite{kristina}.

- The method of Lagrange multipliers has been much used recently
 for non-equilibrium quantum systems \cite{antal,eisler1,kosov,eisler2},
but we should stress that the physical meaning is completely different :
in these quantum studies, the dynamics is governed by some given quantum Hamiltonian,
and the Lagrange multiplier associated with a macroscopic current is added to the Hamiltonian, whereas in the Filyokov-Karpov approach \cite{russes,russeseq,russesneareq,favretti} or in the Evans approach \cite{evans,evansinvariants,evansetal,evansreview}
the dynamical rules are not given at the beginning but on the contrary
they are determined by the optimization of the dynamical entropy of Eq. \ref{straj}
in the presence of appropriate constraints.

The aim of this paper is to 
give a self-contained and unified presentation of this type of approach,
to discuss the various results that can be obtained, and to make the links with 
other recent developments in the field of non-equilibrium statistical physics.

The paper is organized as follows.
In section \ref{sec_general}, we describe how the maximization of the dynamical entropy
of Eq. \ref{straj} leads to  generalized Gibbs distribution for the probability 
distribution ${\cal P}(\Omega_{[0,t]})$ of dynamical trajectories, and to some fluctuation relation of the integrated current.
In the remainder of the paper, we discuss in details how this general formal argument can be given a more precise meaning
when one considers well defined stationary stochastic dynamics.
In section \ref{defmarkov}, we recall the important properties of
 discrete-time Markov chains, that constitute the simplest framework to describe
stochastic dynamics with a time-extensive dynamical entropy
 (Eqs \ref{straj} and \ref{hKS}).
In section \ref{sec_eq}, we explain how the maximization of the dynamical entropy
of Eq. \ref{straj} in the presence of a constraint concerning the energy
of dynamical trajectories 
allows to recover the usual Boltzmann-Gibbs distribution of equilibrium.
In section \ref{sec_noneq}, we add a supplementary constraint concerning
the flux of some observable and derive the form of the optimal solution
for the stationary distribution and for the transition rates.
We  then focus on the problem of finding effective local stochastic rules,
and explain how a consistent formulation can be obtained by the maximization of the relative
dynamical entropy with respect to the equilibrium trajectories :
we describe the case of discrete-time Markov Chains in section \ref{localnoneq} 
and the case of continuous-time Master Equations in section \ref{sec_master}.
In both cases, we make the link with the Evans approach called 'Non-equilibrium Counterpart to detailed balance' \cite{evans,evansinvariants,evansetal,evansreview}.
Our conclusions are summarized in section \ref{sec_conclusion}.
Appendix \ref{static} contains a reminder on the maximization of static entropy
for the equilibrium, to make the comparison with the maximization of the dynamical entropy discussed in the text.
 In Appendix \ref{sec_alter}, we explain the technical simplifications that occur
if one considers an alternate Markov chain as done originally \cite{russes,russeseq,russesneareq,favretti}.

 \section{ General idea : generalized Gibbs distribution for dynamical trajectories } 

\label{sec_general}

As recalled in Appendix \ref{static}, 
the usual Boltzmann-Gibbs distribution for equilibrium configurations can be obtained 
from the maximization of the {\it static} entropy in the presence of constraints introduced via Lagrange multipliers.
This simple derivation of the equilibrium has been introduced by Jaynes \cite{jaynes57},
and can be found nowadays in most textbooks on statistical physics.
Although Jaynes has often proposed to apply the same strategy
to non-equilibrium phenomena, in particular in \cite{jaynes85},
we have not been able to find in his articles
the expression of Eq. \ref{straj} for the dynamical entropy, whereas  
Eq.  \ref{straj} is the clear starting point of some 'maximization'
in the works of Filyokov and Karpov
\cite{russes,russeseq,russesneareq} and in the Evans approach
\cite{evans,evansinvariants,evansetal,evansreview}.
(Note that  Eq.  \ref{straj} is also the starting point in the works of Dewar \cite{dewar}, but with the different 
goal of justifying the 'maximum production entropy principle').

\subsection{ Maximization of the dynamical entropy with constraints }

For non-equilibrium steady states characterized by some macroscopic current $J_0$, the general idea is the following :
one wishes to optimize the dynamical entropy of Eq. \ref{straj}
with constraints concerning the normalization of Eq. \ref{normatraj}
\begin{eqnarray}
{ \cal N} \equiv \sum_{\Omega_{[0,t]}} {\cal P}(\Omega_{[0,t]}) =1
\label{normatrajbis}
\end{eqnarray}
the averaged energy of the trajectory
\begin{eqnarray}
E^{dyn}(t)  \equiv  \sum_{\Omega_{[0,t]}} {\cal P}(\Omega_{[0,t]})  E(\Omega_{[0,t]}) = t E_0
\label{etrajbis}
\end{eqnarray}
and the averaged current of the trajectory
\begin{eqnarray}
J^{dyn}(t)  \equiv  \sum_{\Omega_{[0,t]}} {\cal P}(\Omega_{[0,t]})  J(\Omega_{[0,t]}) = t J_0
\label{jtrajbis}
\end{eqnarray}

The optimization of the Lagrange functional
\begin{eqnarray}
\psi && = S^{dyn}(t) - \rho \left( { \cal N} -1 \right) - \beta \left( E^{dyn}(t)   - t E_0 \right) + \nu \left( J^{dyn}(t)   - t J_0 \right) 
\nonumber \\
&& = - \sum_{\Omega_{[0,t]}}  {\cal P}(\Omega_{[0,t]}) \ln {\cal P}(\Omega_{[0,t]}) 
- \rho \left( \sum_{\Omega_{[0,t]}} {\cal P}(\Omega_{[0,t]})-1 \right)
\nonumber \\
&& 
 - \beta \left( \sum_{\Omega_{[0,t]}} {\cal P}(\Omega_{[0,t]})  E(\Omega_{[0,t]})  - t E_0 \right) + \nu \left( 
 \sum_{\Omega_{[0,t]}} {\cal P}(\Omega_{[0,t]})  J(\Omega_{[0,t]})   - t J_0 \right) 
\label{psitraj}
\end{eqnarray}
with respect to ${\cal P}(\Omega_{[0,t]})$
\begin{eqnarray}
0= \frac{ \delta \Psi }{\delta  {\cal P}(\Omega_{[0,t]}) }\
= -  \ln {\cal P}(\Omega_{[0,t]}) -1
- \rho 
 - \beta  E(\Omega_{[0,t]})  + \nu   J(\Omega_{[0,t]})   
\label{optipsitraj}
\end{eqnarray}
directly leads to the following  generalized Gibbs distribution for dynamical trajectories
\begin{eqnarray}
{\cal P}(\Omega_{[0,t]}) = e^{ -1 - \rho  - \beta  E(\Omega_{[0,t]})  + \nu   J(\Omega_{[0,t]})   }
\label{soluoptipsitraj}
\end{eqnarray}
where the Lagrange multipliers $\rho$, $\beta$ and $\nu$ are fixed respectively by
the constraints of Eqs 
\ref{normatrajbis}, \ref{etrajbis} and \ref{jtrajbis}.

\subsection{ Maximization of the relative dynamical entropy with respect to some equilibrium reference process }

Another formulation that turns out to be more appropriate in many cases consists in considering
the {\it relative} dynamical entropy
with respect to the corresponding known equilibrium distribution $P^{eq} (\Omega_{[0,t]})$
of trajectories in the absence of the current
\begin{eqnarray}
S^{dyn}_{rel} (t) \equiv - \sum_{\Omega_{[0,t]}}
 {\cal P}(\Omega_{[0,t]}) \ln \frac{ {\cal P}(\Omega_{[0,t]}) }{{\cal P}^{eq}(\Omega_{[0,t]})}
\label{strajrel}
\end{eqnarray}
The idea is then to maximize this relative entropy in the presence of the constraints concerning the normalization
of Eq. \ref{normatrajbis} and the imposed current (Eq. \ref{jtrajbis}) to obtain
\begin{eqnarray}
{\cal P}(\Omega_{[0,t]}) = {\cal P}^{eq}(\Omega_{[0,t]}) e^{ -1 - \rho   + \nu   J(\Omega_{[0,t]})   }
\label{soluoptipsitrajrel}
\end{eqnarray}
instead of Eq. \ref{soluoptipsitraj}.

 The notion of relative entropy with respect to some reference, also called the Kullback-Leibler entropy, is well-known in many areas. 
At a naive level, the idea is that the probability distribution
 used as reference constitutes some 'prior' for the problem at hand :
 in the absence of any other constraint, the relative entropy is maximal for the 'prior' distribution. At a more fundamental level, the importance of 
the notion of relative entropy comes from its essential role in the Sanov's theorem 
concerning the theory of large deviations when formulated at the so-called 'level-2'
(see for instance the review \cite{touchette}).

 The interest to consider relative entropies in various areas of statistical physics has been discussed recently in \cite{banavar}.
In the present context of non-equilibrium steady states, the idea to take the equilibrium dynamics as the reference process is the basis of the Evans approach \cite{evans,evansinvariants,evansetal,evansreview}.

 \subsection{ Consequence : fluctuation relation concerning the integrated current } 
 
 \label{fluct_current}

It is clearly impossible to summarize here all the recent developments
concerning the various 'fluctuation relations' that have been established 
in the field of non-equilibrium dynamics,
and we refer the interested reader to some recent reviews 
\cite{derrida,harris_Schu,kurchan,searles,zia,maes,chetrite} 
and references therein.
Here we simply mention the minimum for our present purpose.

The probability distribution of the integrated current is expected to 
satisfy some large deviation principle: 
the probability to have a given time-averaged value $J(\Omega_{[0,t]})/t=j$
behaves at large $t$ as
\begin{eqnarray}
Prob\left( \frac{J(\Omega_{[0,t]})}{t} =j  \right)  \oppropto_{t \to +\infty} 
e^{ t G(j) }
\label{noneqGj}
\end{eqnarray}
where $G(j) \leq 0$ is called the large-deviation function.
The typical value $j_{typ}$ corresponding to the point
where it vanishes $G(j_{typ})=0$ is fixed here to $J_0$ by the constraint of Eq. \ref{jenerj0}.
Equivalently, the generating function of the integrated current $ J(\Omega_{[0,t]})$
behaves for large time $t$ as
\begin{eqnarray}
\langle  e^{\lambda  J(\Omega_{[0,t]})} \rangle  \oppropto_{t \to +\infty}  \int dj e^{ t (\lambda j + G(j) )}
\oppropto_{t \to \infty} e^{t \mu(\lambda)}
\label{noneqmu}
\end{eqnarray}
where $\mu(\lambda)$ is the Legendre transform of $G(j)$
\begin{eqnarray}
 \mu(\lambda) = \opmax_{j} \left[ \lambda j + G(j) \right]
\label{mulambdalegendre}
\end{eqnarray}

To characterize the irreversibility of the dynamics,
it is interesting to compare the probabilities
of a given trajectory $\Omega_{[0,t]}$
and of its associate trajectory 
 ${\tilde \Omega}_{[0,t]}$ obtained by time-reversal. In our present case, 
these two trajectories have the same energy
$E({\tilde \Omega}_{[0,t]})=E(\Omega_{[0,t]}) $ but have
opposite currents $J({\tilde \Omega}_{[0,t]})=- J(\Omega_{[0,t]}) $.
So Eq. \ref{soluoptipsitraj} yields that the ratio of 
the probabilities between a trajectory and its time-reversed trajectory reads
\begin{eqnarray}
\frac{ {\cal P}(\Omega_{[0,t]}) }
{{\cal P}({\tilde \Omega}_{[0,t]}) } 
=  e^{ 2 \nu J(\Omega_{[0,t]})}  
\label{pomegareversed}
\end{eqnarray}
More generally,  any equilibrium dynamics has the property that two reversed-time trajectories
occur with the same probability  $ {\cal P}^{eq}(\Omega_{[0,t]}) =
{\cal P}^{eq}({\tilde \Omega}_{[0,t]}) $. As a consequence, Eq. \ref{soluoptipsitrajrel} also leads to Eq. \ref{pomegareversed}.

A direct consequence of Eq. \ref{pomegareversed}
is that the generating function of the integrated current defined as
\begin{eqnarray}
\langle e^{ \lambda J(\Omega_{[0,t]}) }  \rangle
\equiv \sum_{\Omega_{[0,t]}} e^{ \lambda J(\Omega_{[0,t]})}
 {\cal P}(\Omega_{[0,t]}) 
\label{generating}
\end{eqnarray}
will satisfy the following symmetry (using the antisymmetry $J({\tilde \Omega}_{[0,t]})=- J(\Omega_{[0,t]}) $
and the one-to-one change of variables between $\Omega_{[0,t]}$ and ${\tilde \Omega}_{[0,t]}$)
\begin{eqnarray}
\langle e^{ \lambda J(\Omega_{[0,t]})} \rangle = \sum_{{\tilde \Omega}_{[0,t]} }
e^{  - (\lambda+2 \nu) J({\tilde \Omega}_{[0,t]}) } {\cal P}({\tilde \Omega}_{[0,t]}) 
=  \langle e^{ -(\lambda+2 \nu) J(\Omega_{[0,t]}) }  \rangle
\label{generatingsym}
\end{eqnarray}
that can be rewritten for the function $\mu(\lambda)$ introduced in
Eq. \ref{noneqmu}
\begin{eqnarray}
\mu(\lambda) = \mu(-\lambda- 2 \nu)
\label{musym}
\end{eqnarray}
or equivalently after the Legendre transform of Eq. \ref{mulambdalegendre}
\begin{eqnarray}
G(j)=G(-j)+ 2 \nu j
\label{symGj}
\end{eqnarray}

In summary, the generalized Gibbs distribution of dynamical trajectories of Eq \ref{soluoptipsitraj}
or Eq \ref{soluoptipsitrajrel} directly leads to some fluctuation relation for the integrated current,
as already discussed by Baule and Evans \cite{evansinvariants}.
Independently of the context of the maximization of the dynamical entropy,
the fact that fluctuation relations have actually for origin some generalized Gibbs distribution of space-time
trajectories has been proposed by Maes \cite{maesgibbs}.

 \subsection{ Entropy production } 
 
  \label{entropyprod}

The notion of entropy 'production' is of course subtle.
Since it is expected to measure the 'irreversibility' of the dynamics,
it is natural to relate it to the probabilities
between a trajectory and its reversed-time trajectory 
already introduced in Eq. \ref{pomegareversed} :
a possibility \cite{maesgibbs,maesprod} is thus to
define the entropy production $S^{prod}(\Omega_{[0,t]})$ of a
dynamical trajectory $ \Omega_{[0,t]}$ as the logarithm 
of the ratio of Eq. \ref{pomegareversed}
\begin{eqnarray}
S^{prod}(\Omega_{[0,t]}) \equiv \ln 
\frac{ {\cal P}(\Omega_{[0,t]}) }
{{\cal P}({\tilde \Omega}_{[0,t]}) } 
\label{defsprodtraj}
\end{eqnarray}
Such a definition, which is in agreement with the usual expressions of the averaged 
entropy production derived for Markov chains \cite{schnakenberg,Leb_spo},
has for advantage that the fluctuation relation
for the entropy production
\begin{eqnarray}
\frac{ Prob(S^{prod}) }{ Prob(- S^{prod}) }= e^{S^{prod}}
\label{fluctuationsprodtraj}
\end{eqnarray}
is then a direct consequence of Eq. \ref{defsprodtraj} because
\begin{eqnarray}
Prob(S^{prod}) && \equiv \sum_{\Omega_{[0,t]}} {\cal P}(\Omega_{[0,t]})
\delta \left( S^{prod}- \ln \frac{ {\cal P}(\Omega_{[0,t]}) }
{{\cal P}({\tilde \Omega}_{[0,t]}) } \right)
= \sum_{\tilde \Omega_{[0,t]}} {\cal P}(\tilde \Omega_{[0,t]}) e^{S^{prod}}
\delta \left( S^{prod} + \ln \frac{ {\cal P}(\tilde\Omega_{[0,t]}) }
{{\cal P}( \Omega_{[0,t]}) } \right)  \nonumber \\
&& = e^{S^{prod}} Prob(- S^{prod})
\label{proofsprodtraj}
\end{eqnarray}

For our present case, the definition of Eq. \ref{defsprodtraj} leads
with Eq.  \ref{pomegareversed} to
\begin{eqnarray}
S^{prod}(\Omega_{[0,t]}) \equiv \ln 
\frac{ {\cal P}(\Omega_{[0,t]}) }
{{\cal P}({\tilde \Omega}_{[0,t]}) } 
=  2 \nu J(\Omega_{[0,t]}) 
\label{sprodtraj}
\end{eqnarray}
i.e.  it is simply proportional
to the time-extensive trajectory current $J(\Omega_{[0,t]})$
 that breaks the time asymmetry.

 \subsection{ Discussion } 

In this section, we have described the general idea of maximization of the dynamical entropy with constraints
that leads to generalized Gibbs distributions for dynamical trajectories, and 
to some fluctuation relation for the integrated current.
However, the derivation presented above remains a bit formal, since the space of dynamical trajectories 
has not been precisely defined. In particular, the important notion of 'stationarity' of the dynamics
has been imposed only
implicitly by the time-extensive constraints of Eqs \ref{etrajbis} and \ref{jtrajbis},
but no direct condition of stationarity has been explicitly imposed on $ {\cal P}(\Omega_{[0,t]})$
(and we do not see how one could impose such a stationarity condition at this formal level).
In the following sections, we thus consider well defined stochastic stationary dynamics
generated by discrete-time Markov Chains or by continuous-time Master Equations
to study  the consequences of the maximisation of the dynamical entropy.

 \section{ Dynamical observables for a stationary Markov chain } 

\label{defmarkov}

Discrete-time Markov Chains constitute the simplest formulation of stochastic dynamics
presenting a dynamical entropy (Eq. \ref{straj}) that is extensive in time 
(Eq. \ref{hKS}).
In this section, we introduce the notations that
 will be useful in the remainder of the paper.

\subsection{Discrete-time Markov Chain }

Let us consider a system where possibles microscopic configurations are indexed by $i$,
and where the probability $P_t(i)$ to be in configuration $i$ at time $t$
evolves according to the discrete-time Markov Chain
\begin{eqnarray}
P_{t+1}(j) = \sum_i P_t(i) W_{i \to j}
\label{markov}
\end{eqnarray}
where $W_{i \to j} $ represents the transition probability
from $i$ to $j$ during the unit time interval,
with the following
normalization for each $i$ 
\begin{eqnarray}
 \sum_j  W_{i \to j} =1
\label{markovnormaw}
\end{eqnarray}
The stationary state $P^{st}(i)$ associated with this Markov chain
satisfies
\begin{eqnarray}
P^{st}(i) = \sum_j P^{st}(j) W_{j \to i}
\label{stmarkov}
\end{eqnarray}
and the normalization 
\begin{eqnarray}
\sum_i P^{st}(i) =1
\label{normastmarkov}
\end{eqnarray}

\subsection{Probability of a dynamical trajectory in the stationary regime }

Suppose one starts at time $t=0$ in the stationary state $P^{st}$ of Eq. \ref{stmarkov}.
The probability of a trajectory $\Omega_{[0,t]}=\{i_0,i_1,...,i_t \}$
to be in state $i_0$ at time $t=0$, in state $i_1$ at time $t=1$,
in state $i_2$ at time $t=2$, ... , and in state $i_t$ at time $t$
reads
\begin{eqnarray}
{\cal P}(\Omega_{[0,t]}=\{i_0,i_1,...,i_t \})
= P^{st}(i_0) W_{i_0 \to i_1}  W_{i_1 \to i_2} ... 
 W_{i_{t-1} \to i_t}
\label{pomegamarkov}
\end{eqnarray}
The normalization of Eq. \ref{normatraj}
\begin{eqnarray}
\sum_{i_0} \sum_{i_1} \sum_{i_2} ...\sum_{i_t}
{\cal P}(\Omega_{[0,t]}=\{i_0,i_1,...,i_t \})
= \sum_{i_0} P^{st}(i_0) \sum_{i_1} W_{i_0 \to i_1}  \sum_{i_2} W_{i_1 \to i_2} ... 
\sum_{i_t} W_{i_{t-1} \to i_t}
= 1
\label{pomegamarkovnorma}
\end{eqnarray}
 can be easily checked using Eqs \ref{markovnormaw} and \ref{normastmarkov}.

\subsection{ Dynamical entropy associated with the distribution of trajectories}

Using the normalization and stationarity properties of Eq. \ref{markovnormaw} and Eq. \ref{stmarkov},
the dynamical entropy of Eq. \ref{straj} reads for the distribution ${\cal P}(\Omega_{[0,t]})$
of Eq \ref{pomegamarkov}
\begin{eqnarray}
&& S^{dyn}(t)  = - \sum_{\Omega_{[0,t]}} {\cal P}(\Omega_{[0,t]}) \ln {\cal P}(\Omega_{[0,t]}) \nonumber \\
&& = - \sum_{i_0} \sum_{i_1}  ...\sum_{i_t}
 P^{st}(i_0) W_{i_0 \to i_1}  ... 
 W_{i_{t-1} \to i_t} \left[
\ln \left( P^{st}(i_0) W_{i_0 \to i_1}
... W_{i_{t-2} \to i_{t-1} }\right) +\ln  W_{i_{t-1} \to i_t} \right]
\nonumber \\
&& = - \sum_{i_0} \sum_{i_1}  ...\sum_{i_{t-1}}
 P^{st}(i_0) W_{i_0 \to i_1}   ...
 W_{i_{t-2} \to i_{t-1}}
 \left[\ln \left( P^{st}(i_0) W_{i_0 \to i_1} 
... W_{i_{t-2} \to i_{t-1} } \right)
 + \sum_{i_t} W_{i_{t-1} \to i_t} 
\ln  W_{i_{t-1} \to i_t} \right] \nonumber \\
&& = S^{dyn}(t-1)
- \sum_{i_{t-1}}
 P^{st}(i_{t-1}) 
\sum_{i_t} W_{i_{t-1} \to i_t} 
\ln  W_{i_{t-1} \to i_t} 
\label{strajmarkov}
\end{eqnarray}
By recursion one obtains the following extensive behavior in the time $t$
\begin{eqnarray}
 S^{dyn}(t)  
= - t  \sum_{i} P^{st}(i) \sum_{j} W_{i \to j} \ln  W_{i \to j} 
\label{strajmarkovfinal}
\end{eqnarray}
 so that the entropy rate of Eq. \ref{hKS} is given by the well-known result
(see for instance the books \cite{book_gaspard,koralov_sinai})
\begin{eqnarray}
h_{KS} = \oplim_{t \to + \infty} \frac{S^{dyn}(t)}{t} 
= -   \sum_{i} P^{st}(i) \sum_{j} W_{i \to j} 
\ln  W_{i \to j} 
\label{hKSmarkov}
\end{eqnarray}
The physical meaning of this formula is clear :
it is the average, over the possible configurations $i$ 
distributed with the stationary distribution $P^{st}(i)$,
of the Shannon entropy associated with the transition
probabilities $W_{i \to j}$ normalized with Eq. \ref{markovnormaw}.

\subsection{ Energy associated with trajectories }

In statistical physics, one usually wishes to fix some constraint concerning the energy.
For instance at equilibrium, the Boltzmann-Gibbs distribution can be obtained
from the optimization of the static entropy in the presence of a constraint
fixing the average energy (see Appendix \ref{static} for a short reminder).
In our present dynamical context, it is natural to associate to each trajectory $\Omega_{[0,t]}$
the integrated energy over the time interval 
\begin{eqnarray}
E(\Omega_{[0,t]}=\{i_0,i_1,...,i_t \}) \equiv  E_{i_0}+E_{i_1}+...E_{i_{t-1}}
\label{eomega}
\end{eqnarray}
so that its averaged value over the trajectories $\Omega_{[0,t]}$ 
(using Eqs \ref{markovnormaw} and \ref{stmarkov} )
\begin{eqnarray}
E^{dyn}(t) && \equiv  \sum_{\Omega_{[0,t]}} {\cal P}(\Omega_{[0,t]})  E(\Omega_{[0,t]}) \ = \sum_{i_0} \sum_{i_1}  ...\sum_{i_t}
 P^{st}(i_0) W_{i_0 \to i_1}  ... 
 W_{i_{t-1} \to i_t}  \left[E_{i_0}+E_{i_1}+...E_{i_{t-1}} \right]\nonumber \\
&& 
= \sum_{i_0} P^{st}(i_0) E_{i_0}+\sum_{i_1} P^{st}(i_1) E_{i_1}+... 
+ \sum_{i_{t-1}} P^{st}(i_{t-1}) E_{i_{t-1}}  = t \sum_{i} P^{st}(i) E_{i}
\label{etraj}
\end{eqnarray}
is extensive in time, and the ratio $E^{dyn}(t)/t $
corresponds to the static averaged energy computed from
the stationary distribution as it should.

\subsection{ Flux associated with trajectories }

In the field of non-equilibrium steady states, one is interested in particular in situations 
where a flux or current of some observable is imposed on the system.
It can be a flux of particles, a flux of energy, or a flux of something else
depending on the system under study and on the environment.

In our present framework, we will consider that each transition $i \to j$ is characterized by
some contribution $J_{i \to j}$ to this current, with the antisymmetry property
\begin{eqnarray}
J_{i \to j} = - J_{j \to i}
\label{antisymJ}
\end{eqnarray}
The integrated current $J(\Omega_{[0,t]})$ associated with a trajectory $\Omega_{[0,t]}$
simply reads
\begin{eqnarray}
J(\Omega_{[0,t]}=\{i_0,i_1,...,i_t \}) \equiv J_{i_0 \to i_1} + J_{i_1 \to i_2}
 + ... + J_{i_{t-1} \to i_t} 
\label{jomega}
\end{eqnarray}
so that its averaged value over the trajectories $\Omega_{[0,t]}$ 
(using Eqs \ref{markovnormaw} and \ref{stmarkov} )
\begin{eqnarray}
J^{dyn}(t) && \equiv  \sum_{\Omega_{[0,t]}} {\cal P}(\Omega_{[0,t]})  J(\Omega_{[0,t]}) \ = \sum_{i_0} \sum_{i_1}  ...\sum_{i_t}
 P^{st}(i_0) W_{i_0 \to i_1}  ... 
 W_{i_{t-1} \to i_t}  \left[ J_{i_0 \to i_1} + J_{i_1 \to i_2}
 + ... + J_{i_{t-1} \to i_t}\right]\nonumber \\
&& 
= \sum_{i_0} P^{st}(i_0) \sum_{i_1} W_{i_0 \to i_1} J_{i_0 \to i_1} +... 
+ \sum_{i_{t-1}} P^{st}(i_{t-1}) \sum_{i_t} W_{i_{t-1} \to i_t} J_{i_{t-1} \to i_t} 
\nonumber \\
&& 
 = t \sum_{i} P^{st}(i) \sum_{j} W_{i \to j} J_{i \to j}
\label{jtraj}
\end{eqnarray}
is again extensive in time as expected.

 \section{ Recovering equilibrium from the maximization of the dynamical entropy } 

\label{sec_eq}

The minimal requirement for any approach concerning non-equilibrium steady-states
is of course to be compatible with the theory of the equilibrium.
In this section, we thus describe how the maximization of the dynamical entropy of Eq. \ref{straj}
allows to recover the equilibrium. As a comparison, we recall in Appendix \ref{static}
how the equilibrium is usually obtained via the maximization of the {\it static } entropy.

\subsection{ Lagrange functional taking into account the constraints } 

\label{seceqmaxi}

We consider all possible stationary distributions $P^{st}(i)$ satisfying the normalization 
\begin{eqnarray}
{\cal N} \equiv \sum_i P^{st}(i) =1
\label{normapst}
\end{eqnarray}
and all Markov chains defined by some transition probabilities $W_{i \to
j}$, that satisfy 
the normalization condition of Eq. \ref{markovnormaw} for all $i$
\begin{eqnarray}
 {\cal N}_i \equiv \sum_j  W_{i \to j} =1
\label{nimarkovnorma}
\end{eqnarray}
and the stationarity condition of Eq. \ref{stmarkov} for all $i$
\begin{eqnarray}
\Sigma_i \equiv \sum_j P^{st}(j) W_{j \to i} - P^{st}(i) =0
\label{sistmarkov}
\end{eqnarray}
We also wish to fix the time averaged energy $E^{dyn}(t)/t$ (see Eq. \ref{eaveq}) to some value $E_0$
\begin{eqnarray}
\frac{E^{dyn}(t)}{t} \equiv  \sum_{i} P^{st}(i) E_{i} =E_0
\label{eav}
\end{eqnarray}

To optimize the dynamical entropy $S^{dyn}(t) $ of Eq. \ref{strajmarkovfinal}
in the presence of all these constraints, we introduce 
 Lagrange multipliers $\rho$, $\beta$,
$\lambda_i$ and $\mu_i$ and we consider 
the functional 
\begin{eqnarray}
\Psi && \equiv \frac{S^{dyn}(t)}{t} 
- \rho \left({\cal N}  -1 \right ) 
- \beta \left(\frac{E^{dyn}(t)}{t}-E_0 \right)
- \sum_i \lambda_i \left({\cal N}_i -1 \right)
- \sum_i \mu_i \left(\Sigma_i  \right)
\nonumber \\
&& = -   \sum_{i} P^{st}(i) \sum_{j} W_{i \to j} 
\ln  W_{i \to j} 
- \rho \left(\sum_i P^{st}(i) -1 \right) 
- \beta \left(\sum_i E_i P^{st}(i)-E_0 \right) \nonumber \\
&& - \sum_i \lambda_i \left(\sum_j  W_{i \to j} -1 \right)
- \sum_i 
\mu_i \left(\sum_j P^{st}(j) W_{j \to i} - P^{st}(i)  \right)
\label{psieqmarkovfull}
\end{eqnarray}

\subsection{ Solving the optimization equations } 

We wish to optimize the functional of Eq. \ref{psieqmarkovfull}
both over the stationary probabilities $P^{st}(i)$
\begin{eqnarray}
0= \frac{ \delta \Psi }{\delta P^{st}(i)  }
=  -   \sum_{j} W_{i \to j} 
\ln  W_{i \to j} 
- \rho
- \beta E_i +\mu_i
- \sum_j \mu_j  W_{i \to j} 
\label{optipsieqmarkovpfull}
\end{eqnarray}
and over the transition probabilities $W_{i \to j}$
\begin{eqnarray}
0= \frac{ \delta \Psi }{\delta W_{i \to j}  }
=      P^{st}(i)  
\left[ - \ln  W_{i \to j} - 1 - \mu_j \right] -  \lambda_i  
\label{optipsieqmarkovwfull}
\end{eqnarray}
The Lagrange multipliers have then to be determined by imposing the various constraints.

Equation \ref{optipsieqmarkovwfull} yields
\begin{eqnarray}
  W_{i \to j}=  e^{   - 1 - \mu_j  -  \frac{\lambda_i }{P^{st}(i)}}
\label{optipsieqmarkovsol}
\end{eqnarray}
The normalization constraint of Eq. \ref{nimarkovnorma} for all $i$
\begin{eqnarray}
1 = \sum_j  W_{i \to j} = e^{   - 1  -  \frac{\lambda_i }{P^{st}(i)}} \sum_j   e^{- \mu_j} 
\label{nimarkovnormasol}
\end{eqnarray}
imply that the Lagrange multipliers $\lambda_i$ are given by
\begin{eqnarray}
       \lambda_i  = P^{st}(i)
\ln \left( \sum_j   e^{- \mu_j -1 } \right)
\label{lambdaisol}
\end{eqnarray}
so that the rates of Eq. \ref{optipsieqmarkovsol} can be rewritten as
\begin{eqnarray}
  W_{i \to j}=  \frac{ e^{  - \mu_j}  }{ \sum_k e^{  - \mu_k} }
 \label{optipsieqmarkovsolsimpli}
\end{eqnarray}

The stationarity constraint of Eq. \ref{sistmarkov} yields that for all $j$
\begin{eqnarray}
P^{st}(j) = \sum_i P^{st}(i) W_{i \to j} =
 \frac{   e^{  - \mu_j}  }{ \sum_k e^{  - \mu_k} }
\label{sistmarkovimposed}
\end{eqnarray}
so that the optimum found in Eq. \ref{optipsieqmarkovsolsimpli}
simply corresponds to \cite{russeseq,favretti}
\begin{eqnarray}
  W_{i \to j}= P^{st}(j)
 \label{soluMarkov}
\end{eqnarray}

Eq. \ref{optipsieqmarkovpfull} becomes 
using Eq \ref{sistmarkovimposed} and Eq \ref{normapst}
\begin{eqnarray}
0= - \rho - \beta E_i - \ln P^{st}(i)
\label{optipsieqmarkovpfullbis}
\end{eqnarray}
so that using the normalization constraint of Eq. \ref{normapst}
one recovers the Boltzmann-Gibbs distribution
\begin{eqnarray}
P^{st}(j) = \frac{ e^{- \beta E_j}}{ \sum_i e^{- \beta E_i} } \equiv P^{eq}(j)
\label{Boltzmann-Gibbsbis}
\end{eqnarray}
where $\beta$ is fixed by the constraint of Eq. \ref{eav}.

\subsection{ Consequence for the probabilities of dynamical trajectories }

With the optimal solution obtained in Eqs \ref{soluMarkov} and \ref{Boltzmann-Gibbsbis},
the probability of a trajectory $\Omega_{[0,t]}$ of Eq. \ref{pomegamarkov}
 takes the simple form
\begin{eqnarray}
{\cal P}^{eq}(\Omega_{[0,t]})=\{i_0,i_1,...,i_t \})
= P^{st}(i_0) W_{i_0 \to i_1}  W_{i_1 \to i_2} ... 
 W_{i_{t-1} \to i_t} = \frac{  e^{- \beta E(i_0) } \ e^{- \beta E(i_1) } ...
\ e^{- \beta E(i_t) } }{ \left[ Z(\beta) \right]^{t+1} }
\label{pomegamarkoveq}
\end{eqnarray}
where
\begin{eqnarray}
Z(\beta) \equiv \sum_i e^{- \beta E_i}
\label{zeq}
\end{eqnarray}
is the usual static partition function.
So within the present approach, 
the probability distribution over dynamical trajectories is simply
the Boltzmann-Gibbs distribution with respect 
to the trajectory energy $E(\Omega_{[0,t+1]})$ of Eq. \ref{etraj}
\begin{eqnarray}
{\cal P}^{eq}(\Omega_{[0,t]}) = 
\frac{ e^{- \beta E(\Omega_{[0,t+1]})} }{ \left[ Z(\beta) \right]^{t+1} }
\label{peqtraj}
\end{eqnarray}
The appearance of $E(\Omega_{[0,t+1]})$ (instead of $E(\Omega_{[0,t]})$ found in Eq. \ref{soluoptipsitraj} via the formal argument ) is a consequence of the choice of
Eq. \ref{etraj} for the energy for the discrete Markov chain : the energy
$E(i_t)$ of the last state $i_t$ will contribute to the trajectory 
energy only during the next time interval $[t,t+1]$, but it nevertheless appear
in the probability of the trajectory $ \Omega_{[0,t]}$ ending in state $i_t$. 
This slight problem thus comes from the discrete nature of the dynamics,
and is not expected to play an important role :  it is only a finite boundary term that will become subleading in the large time limit $t \to +\infty$
with respect to the time-extensive energy of the whole trajectory.

An essential property of this distribution of trajectories
is the symmetry by time-reversal : the probabilities
of a given trajectory $\Omega_{[0,t]}=\{i_0,i_1,...,i_{t-1},i_t\}$
and of its associate trajectory 
 ${\tilde \Omega}_{[0,t]}=\{i_t,i_{t-1},...,i_{1},i_0\}$
obtained by time-reversal are equal
\begin{eqnarray}
\frac{ {\cal P}(\Omega_{[0,t]}=\{i_0,i_1,...,i_t \}) }
{{\cal P}({\tilde \Omega}_{[0,t]}=\{i_t,i_{t-1},...,i_{1},i_0\}) } 
= 1 
\label{eqreversed}
\end{eqnarray}

\subsection{ Discussion } 

\label{discussion_eq}

In this section, we have described how the maximization of the dynamical entropy allows
to recover that the stationary distribution follows the Boltzmann-Gibbs distribution (Eq. \ref{Boltzmann-Gibbsbis}). 
However the result of Eq. \ref{soluMarkov} concerning the transition probabilities
 $W_{i \to j}$ may be surprising at first sight :  
this solution means that the new configuration $j$ 
is chosen with the probability $P^{st}(j)$
 and is completely independent of the initial state $i$. 
Physically this corresponds to a very coarse-grained dynamics with no memory.
It is thus clear that  Eq. \ref{soluMarkov} does not represent an effective
'microscopic' dynamics, but represents an effective dynamics on some macroscopic time $\tau$. The corresponding 
entropy rate per unit time of Eq. \ref{hKS} 
\begin{eqnarray}
h_{KS} 
= -   \left( \sum_{i} P^{st}(i) \right) \sum_{j} P^{st}(j) \ln  P^{st}(j) 
= -    \sum_{j} P^{st}(j) \ln  P^{st}(j) 
\label{hKSmarkovopt}
\end{eqnarray}
then exactly coincides with the static entropy of Eq. \ref{seq},
which seems natural if one wishes this approach to be equivalent to
the maximization of the static entropy that one uses for the equilibrium
 (see Appendix \ref{static}).
This discussion suggests the following interpretation of
 this type of computation.

\subsection{ Final formulation of the physical meaning of this approach } 

\label{meaningeq}

The statistical physics theory of equilibrium is usually based
on some 'ergodic' hypothesis, stating that time-averages of 
 observables $A(i)$ will converge in the infinite-time limit
towards averages computed with respect 
to the Boltzmann-Gibbs distribution $P^{eq}(i)$ of Eq. \ref{Boltzmann-Gibbsbis}
\begin{eqnarray}
\lim_{t \to +\infty} \frac{1}{t} \int_0^t dt A(i(t)) = \sum_i A(i) P^{eq}(i)
\label{ergodic}
\end{eqnarray}
Since the right-handside contains only the Boltzmann-Gibbs distribution $P^{eq}(i)=e^{-\beta E_i}/Z(\beta)$ and no other information about the dynamics
except the conserved energy, this means that the precise form of the 'true'
microscopic dynamics becomes completely irrelevant at large time.
Loosely speaking, this means that there exists some finite macroscopic 
correlation time $\tau_{correl}$, beyond which the dynamical correlations
have been lost in practice, so that the large-time interval $t$ can be decomposed into $t/\tau_{correl}$ quasi-independent time intervals, and
  Eq. \ref{ergodic} becomes possible. On the contrary, if the correlation time $\tau_{correl}$ of the 'true' microscopic dynamics is infinite, Eq. \ref{ergodic} cannot
really be satisfied since other dynamical informations besides
the conservation of energy remain relevant forever.

This discussion suggests that the analysis presented above based on the maximization of the dynamical entropy, with the result of Eq. \ref{soluMarkov},
actually describes what happens at this coarse-grained macroscopic
scale $\tau_{correl}$, i.e. the unit-time of the Markov Chain of Eq. \ref{markov} should be interpreted as $\tau_{correl}$.
Then we have shown above that one recovers the Boltzmann-Gibbs distribution,
and thus all the 'static' properties that can be derived from it.

However in the statistical physics of equilibrium, once one has understood
the properties of the Boltzmann-Gibbs measure over configurations,
one can become interested into the 'equilibrium dynamics'.
However, one does not wish to return to the 'true' deterministic microscopic dynamics which is usually very complicated for systems with 
a very large number (like $10^{23}$) degrees of freedom, because
all the details of the dynamics
 have proven to be irrelevant at large time scale.
So one introduces an {\it effective stochastic microscopic dynamics }
that is compatible with the known properties on large time scales.
For definiteness, let us consider a {\it microscopic} Markov Chain
\begin{eqnarray}
p_{t+\Delta t}(j) = \sum_i p_t(i) w_{i \to j}
\label{markovmicro}
\end{eqnarray}
where $w_{i \to j} $ represents the transition probability
from $i$ to $j$ during the microscopic time interval $\Delta t$,
so that $w_{i \to j} $  are now non-zero only if
the configurations $i$ and $j$ are sufficiently close in configuration space.
For instance in a system of $N$ spins with $2^N$ configurations, 
one may require that $w_{i \to j} $ is non-zero only if $j$ can be obtained from $i$ by the flip a single spin. 
More generally, this notion of elementary moves has to be
defined in an appropriate way for each type of models.
Then the question is : beyond this locality requirement, how should
these microscopic transition probabilities $w_{i \to j} $ be chosen
to be compatible with the properties known on large time scales ?
The first obvious requirement is that the microscopic Markov Chain of Eq.
\ref{markovmicro} should have for stationary state $p^{st}(i)$ 
the Boltzmann-Gibbs distribution which is known to be the stationary state
on macroscopic times
\begin{eqnarray}
p^{st}(i) = P^{eq}(i)
\label{pstmicro}
\end{eqnarray}
However this is not the only constraint, since one also wishes
to reproduce the essential time-reversibility property of dynamical trajectories of Eq. \ref{eqreversed}. For this, it is sufficient to impose that
the probability of the elementary 
microscopic trajectory $\omega_{\Delta t}=\{i_0,i_1\}$
\begin{eqnarray}
{Prob}(\omega_{\Delta t}=\{i_0,i_1\})=p(i_0) w_{i_0 \to i_1} 
\label{psttrajmicro}
\end{eqnarray}
is equal to the probability of the reversed trajectory ${\tilde \omega}_{\Delta t}=\{i_1,,i_0\} $
\begin{eqnarray}
1 = \frac{ {Prob(\omega_{\Delta t}=\{i_0,i_1\}) } }
{ { Prob }({\tilde \omega}_{\Delta t}=\{i_1,,i_0\}) } 
= \frac{p(i_0) w_{i_0 \to i_1} }{p(i_1) w_{i_1 \to i_0}} 
\label{eqreversedmicro}
\end{eqnarray}
Taking into account Eq. \ref{pstmicro}, this leads to the well-known
detailed-balance condition
\begin{eqnarray}
\frac{ w^{eq}_{i_0 \to i_1} }{ w^{eq}_{i_1 \to i_0}} 
= \frac{p(i_1)  }{p(i_0)} =  \frac{ P^{eq}(i_1) }{P^{eq}(i_0)}= e^{-\beta(E(i_1)-E(i_0))}
\label{detailed}
\end{eqnarray}
Besides this detailed-balance constraint, there is still some freedom
to choose the effective microscopic transition probabilities $w_{i \to j} $,
but one usually considers that they are equally valid, in the sense that they will
all reproduce the essential large-time properties of convergence
towards the Boltzmann-Gibbs distribution while preserving the time-reversibility
of dynamical trajectories.

\subsection{ Is it possible to add locality constraints within the maximization procedure ?  } 

\label{localeq}

As explained above, the requirements on effective local Markov chain dynamics of Eq. \ref{markovmicro}
are derived from the macroscopic equilibrium properties. A natural question here is whether these requirements
can be instead directly derived from a maximisation of the dynamical entropy in the presence of locality constraints on the $w_{i \to j} $.
We have now the following picture : configurations are the nodes of a network, and the links are present
between configurations that are related by an elementary local move.
As a first example,  in a spin models with $N$ spins and $2^N$ configurations, one may consider a single spin-flip dynamics,
where each configuration has $N$ neighbors  (corresponding to the flip of one of its $N$ spins).
As a second example, in lattice gases models with hard-core interactions, the number of neighbors of a configuration
will be given by the number of possible local moves from this configuration, and will thus depend on the configuration.

Whenever the connectivities (i.e. the numbers of neighbors) are configuration-dependent,
it is clear that one cannot recover the equilibrium from the maximization of the dynamical entropy in the presence of locality constraints,
as can be easily understood on the simple infinite temperature case $\beta=0$ :

(i) At infinite temperature, the problem of the maximization of the dynamical entropy
is completely equivalent to the Maximal Entropy Random Walk (MERW) on arbitrary networks studied in details in \cite{jmluck},
with the following main conclusion : in networks where the connectivities of the various nodes
are not all the same, the MERW tends to visit more the sites with higher connectivities than the sites with lower connectivities,
because sites with higher connectivities are associated with a bigger number of choices  for the next step of the random walk.

(ii) On the other hand, at infinite temperature case $\beta=0$, the Boltzmann-Gibbs distribution simply corresponds
to the uniform distribution over all configurations,
independently of their connectivities from the point of view of the local dynamics.

Our conclusion is thus that the choice of the connectivities that one wishes to impose on the local dynamics
has a too strong effect on the dynamical entropy that one wishes to maximize, whereas 
the maximization of the dynamical entropy at a macroscopic time scale allows to recover
the equilibrium, as explained above in section \ref{meaningeq}.
In the non-equilibrium case, this discussion on local dynamics will be the subject of sections \ref{localnoneq} and \ref{sec_master},
but in the next section, we first describe what happens in the absence of locality constraints.

 \section{ Non-equilibrium steady state with an imposed current  } 

\label{sec_noneq}

\subsection{ Lagrange functional taking into account the constraints }

With respect to the previous section \ref{seceqmaxi}, we now add another constraint concerning the flux of Eq. \ref{jtraj} 
\begin{eqnarray}
\frac{J^{dyn}(t)}{t} \equiv \sum_{i} P^{st}(i) \sum_{j} W_{i \to j} J_{i \to j} =J_0
\label{jenerj0}
\end{eqnarray}
So we introduce this constraint with a new Lagrange multiplier $\nu$ into the functional of Eq. \ref{psieqmarkovfull}
to obtain
\begin{eqnarray}
\Psi && \equiv \frac{S^{dyn}(t)}{t} 
- \rho({\cal N} -1) 
- \beta(\frac{E^{dyn}(t)}{t}-E_0 )
- \sum_i \lambda_i \left({\cal N}_i -1 \right)
- \sum_i \mu_i \left(\Sigma_i  \right)
+ \nu \left( \frac{J^{dyn}(t)}{t} - J_0  \right)
\nonumber \\
&& = -   \sum_{i} P^{st}(i) \sum_{j} W_{i \to j} 
\ln  W_{i \to j} 
- \rho(\sum_i P^{st}(i) -1) 
- \beta(\sum_i E_i P^{st}(i)-E_0 ) \nonumber \\
&& - \sum_i \lambda_i \left(\sum_j  W_{i \to j} -1 \right)
- \sum_i 
\mu_i \left(\sum_j P^{st}(j) W_{j \to i} - P^{st}(i)  \right)
+ \nu \left( \sum_{i} P^{st}(i) \sum_{j} W_{i \to j} J_{i \to j} - J_0  \right)
\label{psimarkovfull}
\end{eqnarray}

\subsection{ Solving  the optimization equations } 

Equations \ref{optipsieqmarkovpfull} and \ref{optipsieqmarkovwfull} are modified into
\begin{eqnarray}
0= \frac{ \delta \Psi }{\delta P^{st}(i)  }
=  -   \sum_{j} W_{i \to j} 
\ln  W_{i \to j} 
- \rho
- \beta E_i +\mu_i
- \sum_k \mu_k  W_{i \to k} 
+   \nu  \sum_{j } W_{i \to j}  J_{i \to j}
\label{optipsimarkovpfull}
\end{eqnarray}
and
\begin{eqnarray}
0= \frac{ \delta \Psi }{\delta W_{i \to j}  }
=      P^{st}(i)  
\left[ - \ln  W_{i \to j} - 1 - \mu_j  + \nu J_{i \to j}  \right] -  \lambda_i 
 \label{optipsimarkovwfull}
\end{eqnarray}

Equation \ref{optipsimarkovwfull} yields
\begin{eqnarray}
  W_{i \to j}=  e^{   - 1 - \mu_j + \nu  J_{i \to j} -  \frac{\lambda_i }{P^{st}(i)}}
\label{optipsimarkovsol}
\end{eqnarray}
The normalization constraint of Eq. \ref{nimarkovnorma} for all $i$
\begin{eqnarray}
1 = \sum_j  W_{i \to j} = e^{   - 1  -  \frac{\lambda_i }{P^{st}(i)}} \sum_j 
  e^{- \mu_j +  \nu J_{i \to j}  } 
\label{nimarkovnormasolnoneq}
\end{eqnarray}
leads to the Lagrange multipliers
\begin{eqnarray}
       \lambda_i  = P^{st}(i)
\ln \left( \sum_j   e^{- \mu_j -1 +  \nu J_{i \to j} } \right)
\label{lambdaisolnoneq}
\end{eqnarray}
so that the rates of Eq. \ref{optipsieqmarkovsol} become
\begin{eqnarray}
  W_{i \to j}=  \frac{ e^{  - \mu_j +  \nu J_{i \to j}  } }
{ \sum_k e^{  - \mu_k +   \nu J_{i \to k}   } }
 \label{optipsimarkovsolsimpli}
\end{eqnarray}

To analyze the stationarity constraint of Eq. \ref{sistmarkov} for all $j$
\begin{eqnarray}
P^{st}(j) = \sum_i P^{st}(i) W_{i \to j} =
\sum_i P^{st}(i) \frac{ e^{  - \mu_j + \nu J_{i \to j} }  }
{ \sum_k e^{  - \mu_k+ \nu J_{i \to k}  }  }
\label{sistmarkovimposednoneq}
\end{eqnarray}
it is convenient to introduce the notations
\begin{eqnarray}
z_i && \equiv \sum_k e^{  - \mu_k  + \nu J_{i \to k}   }
\label{zi}
\end{eqnarray}
and
\begin{eqnarray}
y_j  \equiv 
\sum_i \frac{ P^{st}(i)}{z_i}   e^{ \nu J_{i \to j}  }
\label{yi}
\end{eqnarray}
to rewrite Eq. \ref{sistmarkovimposednoneq} as
\begin{eqnarray}
e^{- \mu_j} = \frac{P^{st}(j)}{y_j} 
\label{sistmarkovimposednoneqbis}
\end{eqnarray}
so we may now eliminate all $\mu_j$ in terms of the $y_j$. In particular
\begin{eqnarray}
z_i = \sum_k \frac{P^{st}(k)}{y_k}  e^{ \nu J_{i \to k}   }  
\label{ziyk}
\end{eqnarray}
is somewhat the 'dual' of Eq. \ref{yi}.
The transition probability of Eq. \ref{optipsimarkovsolsimpli} now reads
\begin{eqnarray}
  W_{i \to j}=  \frac{P^{st}(j)}{z_i y_j }  e^{ \nu J_{i \to j}   }  
 \label{optipsieqmarkovsolsimplinoneqbis}
\end{eqnarray}
and 
Eq. \ref{optipsimarkovpfull} becomes
\begin{eqnarray}
0
&& = - \rho
- \beta E_i - \ln P^{st}(i) + \ln (y_i)
+   \sum_{j} W_{i \to j} \left[ -  \ln  W_{i \to j} - \mu_j + 
\nu J_{i \to j}  \right] \nonumber \\
&& = - \rho
- \beta E_i - \ln P^{st}(i) + \ln (y_i) + \ln(z_i) 
\label{optipsimarkovpfullbis}
\end{eqnarray}
leading to
\begin{eqnarray}
 P^{st}(i) = y_i z_i e^{ - \rho - \beta E_i } 
\label{pstnoneq}
\end{eqnarray}
Plugging Eq. \ref{pstnoneq} into Eqs \ref{yi} and \ref{ziyk} leads to
\begin{eqnarray}
y_j  = 
\sum_i y_i e^{ - \rho - \beta E_i }    e^{ \nu J_{i \to j}   }
\label{yibis}
\end{eqnarray}
and
\begin{eqnarray}
z_i = \sum_k  z_k e^{ - \rho - \beta E_k }  e^{ \nu J_{i \to k}   }  
\label{ziykbis}
\end{eqnarray}

\subsection{ Optimal stationary distribution $P^{st}(i)$ and transition probabilities $W_{i \to j} $ } 

In summary, we have obtained that the 
optimal stationary distribution $P^{st}(i)$ and transition probabilities $W_{i \to j} $
follow the form 
\begin{eqnarray}
 P^{st}(i) && =  y_i z_i e^{ -\rho - \beta E_i } \nonumber \\
 W_{i \to j} && =    \frac{ z_j }{ z_i  } 
 e^{ -\rho  - \beta E_j + \nu J_{i \to j} }  
\label{pstwnoneq}
\end{eqnarray}
where the $y_i$ and $z_i$ are positive variables satisfying 
respectively the equations
\begin{eqnarray}
y_j  = \sum_i y_i e^{ - \rho - \beta E_i }    e^{ \nu J_{i \to j}   }
\label{yiter}
\end{eqnarray}
that correspond to the stationarity constraints $P^{st}(j)=\sum_i P^{st}(i)
W_{i \to j} $
and the equations
\begin{eqnarray}
z_i = \sum_j    e^{ \nu J_{i \to j}   }  e^{ - \rho - \beta E_j }z_j 
\label{ziter}
\end{eqnarray}
that correspond to the normalizations $1=\sum_j W_{i \to j} $,
and 
\begin{eqnarray}
 1  =   \sum_i y_i z_i e^{-\rho  - \beta E_i }  
\label{rhofin}
\end{eqnarray}
that correspond to the normalization $1=\sum_i P^{st}(i)$.
To see more clearly what this structure means,
it is convenient to introduce the bra and ket notations and to set
\begin{eqnarray}
y_i  \equiv   e^{  \frac{\beta}{2} E_i }  <L \vert i> \nonumber \\
z_i \equiv   e^{  \frac{\beta}{2} E_i }  <i \vert R>
\label{leftright}
\end{eqnarray}
Then in terms of the matrix $M$ defined by the positive matrix elements 
\begin{eqnarray}
<i \vert M \vert j> \equiv  e^{  - \frac{\beta}{2} E_i }
e^{ \nu J_{i \to j}   } e^{  - \frac{\beta}{2} E_j }
\label{matrixM}
\end{eqnarray}
Eqs \ref{yiter} and \ref{ziter} mean that $<L \vert$ and $\vert R>$
are respectively positive left-eigenvector and right-eigenvector 
of the non-symmetric matrix $M$
\begin{eqnarray}
 e^{\rho}  <L \vert   &&= <L \vert M \vert
 \nonumber \\
 e^{\rho}  \vert R> && = M \vert R>
\label{eigenLR}
\end{eqnarray}
associated with the highest eigenvalue $e^{\rho}$ of the matrix $M$ 
(Perron-Frobenius), with the normalization given by Eq \ref{rhofin}
\begin{eqnarray}
 e^{\rho}  =   <L  \vert R>  
\label{normaLR}
\end{eqnarray}

The equilibrium case discussed in the previous section \ref{sec_eq} corresponds to $\nu=0$, $y_i=1=z_i$, $<L \vert i>=<i \vert R>= e^{  -\frac{\beta}{2} E_i }$. In the following to discuss what happens for $\nu>0$, we will use
the notations $y_i$ and $z_i$ that appear in Eqs \ref{pstwnoneq}

\subsection{ Consequence for the probability distribution of dynamical trajectories } 

\label{trajboundaryterm}

From the optimal solution of Eq \ref{pstwnoneq}, one obtains that
the probability of a trajectory $\Omega_{[0,t]}$ of Eq. \ref{pomegamarkov}
 takes the form
\begin{eqnarray}
{\cal P}(\Omega_{[0,t]})=\{i_0,i_1,...,i_t \})
&& = P^{st}(i_0) W_{i_0 \to i_1}  W_{i_1 \to i_2} ... 
 W_{i_{t-1} \to i_t}\nonumber \\
&& =  y_{i_0} z_{i_0} e^{ -\rho - \beta E_{i_0} }
 \frac{ z_{i_1} }{ z_{i_0}  } 
 e^{ -\rho  - \beta E_{i_1} + \nu J_{i_0 \to i_1} }
 \frac{ z_{i_2} }{ z_{i_1}  } 
 e^{ -\rho  - \beta E_{i_2} + \nu J_{i_1 \to i_2} } ... 
 \frac{ z_{i_{t}} }{ z_{i_{t-1}}  } 
 e^{ -\rho  - \beta E_{i_t} + \nu J_{i_{t-1} \to i_t} } \nonumber \\
&& = y_{i_0} z_{i_t} e^{ - \beta \left(E_{i_0}+ E_{i_1} + ... + E_{i_t} \right)
 + \nu \left(J_{i_0 \to i_1}+ J_{i_1 \to i_2} +...+ J_{i_{t-1} \to i_t}  \right) } e^{- (t+1)\rho} \nonumber \\
&& = y_{i_0} z_{i_t}  e^{ - \beta E(\Omega_{[0,t+1]})+ \nu J(\Omega_{[0,t]})} e^{- (t+1)\rho} 
\label{pomegamarkovNONEQeq}
\end{eqnarray}
So it is not exactly as simple as Eq. \ref{soluoptipsitraj} :
besides the expected Boltzmann-Gibbs factors involving the trajectory energy $
E(\Omega_{[0,t+1]})$ (already found for the equilibrium case in Eq. \ref{peqtraj})
and the trajectory current $ J(\Omega_{[0,t]})$,
and besides the normalization insured by the choice of $\rho$ in the last factor,
there remains the non-trivial boundary prefactor $ y_{i_0} z_{i_t}$ : 
we believe that this term comes from the stationary constraint on the dynamics that we could impose within the Markov chain framework,
whereas we were not able to impose this stationary constraint at the formal level discussed in section \ref{sec_general}.
However, this supplementary prefactor is a 'boundary term' containing only the initial and the final configurations.
As a consequence, it is not expected to grow in time, and in the large time limit
$t \to +\infty$, it will become subleading  with respect to the time-extensive terms present in the exponential, since the trajectory energy and the trajectory 
 current are extensive in time by the imposed constraints.
(Note however that in some cases with unbounded phase space, the 'boundary terms'
may diverge and remain important even in the large-time limit, see section 5.3 of the review \cite{harris_Schu} and references therein for more details).

 Our conclusion is thus that the formal solution of Eq. \ref{soluoptipsitraj} 
neglects only the boundary terms of Eq. \ref{pomegamarkovNONEQeq}
and thus captures correctly the dominant terms in the large-time limit $t \to + \infty$.
In particular, it contains the factor that is responsible for the fluctuation relation
of the integrated current discussed in section \ref{fluct_current}
and for the dominant term of the entropy production discussed in section \ref{entropyprod}.
We will obtain the same conclusion in the presence of locality constraints on the dynamics
that we consider in the next section.

\section{ Effective stochastic microscopic dynamics in the non-equilibrium case  } 

\label{localnoneq}

As explained in section \ref{localeq}, there exists some difficulty to recover the equilibrium
if one tries to maximize the dynamical entropy associated with a Markov Chain containing locality constraints,
as a consequence of the possible configuration-dependent connectivities that one wishes to impose on the 
possible elementary moves. 
However, as recalled in section \ref{meaningeq}, the requirements on effective stochastic microscopic dynamics 
(Eq. \ref{markovmicro})
to describe the equilibrium dynamics are well understood, and lead to the detailed balance condition of Eq. \ref{detailed}.
In this section, we will thus follow the point of view of the Evans approach \cite{evans,evansinvariants,evansetal,evansreview} :
we consider that a local equilibrium dynamics generated by some Markov Chain $w^{eq}_{i \to j}$ satisfying detailed balance
is known, and we wish to determine the non-equilibrium appropriate modified Markov Chain $w_{i \to j}$
in the presence of an imposed current. Instead of maximizing the 'full' dynamical entropy of Eq. \ref{straj},
we will thus maximize the relative dynamical entropy with respect to the equilibrium dynamics,
that we have introduced in Eq. \ref{strajrel}.

\subsection{ Maximization of the relative dynamical entropy with respect to the equilibrium  } 

For the microscopic Markov Chain
\begin{eqnarray}
p_{t+\Delta t}(j) = \sum_i p_t(i) w_{i \to j}
\label{markovmicrononeq}
\end{eqnarray}
the relative dynamical entropy of Eq. \ref{strajrel} with respect to the equilibrium dynamical trajectory corresponding to
$w^{eq}_{i \to j}$  has for time-extensive behavior 
\begin{eqnarray}
 S^{dyn}_{rel} (t) \equiv - \sum_{\Omega_{[0,t]}}
 {\cal P}(\Omega_{[0,t]}) \ln \frac{ {\cal P}(\Omega_{[0,t]}) }{{\cal P}^{eq}(\Omega_{[0,t]})} 
 \opsimeq_{ t \to + \infty}  - \frac{t }{\Delta t}  \sum_i p^{st}(i) \sum_j 
 w_{i \to j} \ln \left( \frac{w_{i \to j}}{w^{eq}_{i \to j}} \right)
\label{strajrelmarkov}
\end{eqnarray}

The maximization of this relative entropy in the presence of the constraints concerning the normalizations
\begin{eqnarray}
 \sum_i p^{st}(i) && =1 \\ \nonumber
 \sum_j  w_{i \to j}  && =1
 \label{normamicro}
\end{eqnarray}
the stationarity condition 
\begin{eqnarray}
 \sum_j p^{st}(j) w_{j \to i} - p^{st}(i) =0
\label{statiomicro}
\end{eqnarray}
and the current
\begin{eqnarray}
\frac{J^{dyn}(t)}{t} \equiv  \frac{1 }{\Delta t} \sum_{i} p^{st}(i) \sum_j w_{i \to j}  J_{i \to j} =J_0
\label{javmicro}
\end{eqnarray}
can be done with Lagrange multipliers, with steps similar to the ones detailed in the previous section.

To write the final result, it is convenient to introduce
the positive non-symmetric matrix
\begin{eqnarray}
<i \vert M \vert j > = w^{eq}_{i \to j} e^{\nu J_{i \to j} } 
\label{solumicroM}
\end{eqnarray}
its largest eigenvalue $e^{ \rho}$, and the corresponding right and left positive eigenvectors $\vert R>$ and $< L \vert$
\begin{eqnarray}
M \vert R> && = e^{ \rho} \vert R> \nonumber \\
<L \vert M && = e^{ \rho} <L \vert
\label{solumicroRL}
\end{eqnarray}
normalized with
\begin{eqnarray}
<L \vert R> = e^{ \rho} 
\label{solumicronormaRL}
\end{eqnarray}
With these notations, the optimal solution reads
\begin{eqnarray}
p^{st}(i) && = e^{- \rho} <i \vert R> <L \vert i > \nonumber \\
w_{i \to j} && = w^{eq}_{i \to j} e^{- \rho +\nu J_{i \to j} } \frac{<j \vert R>}{<i \vert R>} = e^{- \rho} <i \vert M \vert j >  \frac{<j \vert R>}{<i \vert R>}
\label{solumicro}
\end{eqnarray}
In the equilibrium case $\nu=0$, the matrix $M$   reduces to the generator of the equilibrium Markov Chain $M_{ij} = w^{eq}_{i \to j}$,
 so the maximal eigenvalue corresponds to $ e^{ \rho_{eq} }=1$,
the right and the left eigenvectors are simply $<i \vert R_{eq}>=1$ and $ <L_{eq} \vert i >=p^{eq}(i) $.

\subsection{ Consequences for the probability distribution of dynamical trajectories } 

From the solution of Eq. \ref{solumicro}, one obtains that
the probability of a trajectory $\omega=\{i_0,i_1,...,i_n \})$ 
 takes the form
\begin{eqnarray}
{\cal P}(\omega=\{i_0,i_1,...,i_n \})
&& = p^{st}(i_0) w_{i_0 \to i_1}  w_{i_1 \to i_2} ... w_{i_{n-1} \to i_n}   \nonumber \\
&& =  e^{- \rho} <i_0 \vert R> <L \vert i _0> w^{eq}_{i_0 \to i_1} e^{- \rho +\nu J_{i_0\to i_1} } \frac{<i_1 \vert R>}{<i_0 \vert R>} ...
w^{eq}_{i_{n-1} \to i_n} e^{- \rho +\nu J_{i_{n-1}\to i_n} } \frac{<i_n \vert R>}{<i_{n-1} \vert R>}
 \nonumber \\
&& = \left[  \frac{<L \vert i _0><i_n \vert R>}{p^{eq}(i_0)} \right]  {\cal P}^{eq} (\omega=\{i_0,i_1,...,i_n \}) e^{  \nu J(\omega)} e^{- (n+1)\rho} 
\label{pomegamarkovNONEQeqmicro}
\end{eqnarray}
Again, it is not exactly as simple as Eq. \ref{soluoptipsitrajrel}:
besides the equilibrium trajectory probability, the expected Boltzmann-Gibbs factors involving  the trajectory current $ J(\Omega_{[0,t]})$,
and besides the normalization insured by the choice of $\rho$ in the last factor,
there remains a  boundary prefactor that depends only on the initial and final configurations.
But this boundary factor is expected to become subleading in the large time limit
$t \to +\infty$, as already discussed in section \ref{trajboundaryterm}.

 \subsection{ Link with the Bayesian approach of Evans \cite{evans} } 

In \cite{evans}, Evans has introduced an approach 
called 'Non-equilibrium Counterpart to detailed balance'. 
The general idea is clearly the same as in the present article, namely
the maximization of the dynamical entropy in the presence of the appropriate constraints.
However, the way of reasoning is slightly different and thus gives other useful points of view :
 Bayes theorem is used to analyse the properties of an elementary trajectory
 belonging to a macroscopic trajectory satisfying the flux constraint with respect to the equilibrium dynamics.
 The outcome is that  the transitions $w_{i \to j}^{driven}$ in the driven case
should have the following form (see Eq. (24) of \cite{evans})
\begin{eqnarray}
w_{i \to j}^{driven} = 
w_{i \to j}^{eq} e^{ \nu J_{i \to j} \Delta t + q_j(\nu)-q_i(\nu)
- Q(\nu) \Delta t }
\label{wdrivenevans}
\end{eqnarray}
From the comparison with Eq. \ref{solumicro}, one obtains the following correspondence :
$Q(\nu) \Delta t$ corresponds to the normalization factor $\rho$, 
whereas $q_j(\nu)$ corresponds to $\ln <j \vert R>$.
The physical interpretation proposed by Evans is that the factor 
$q_j(\nu)$ measures the 'willingness' of state $j$ to accept a future flux.
We refer the reader to the very interesting series of articles 
\cite{evans,evansinvariants,evansetal,evansreview} to have more detailed explanations
and to see various examples of application.

\section{Case of continuous-time master equation}

\label{sec_master}

Up to now we have considered the case of discrete-time Markov Chains,
that constitute the simplest framework to define probabilities of 
dynamical trajectories and their Shannon entropy of Eq. \ref{straj}.
However, many studies on non-equilibrium systems prefer to
consider stochastic dynamics that are generated by
some continuous-time Master Equation
\begin{eqnarray}
\partial_t p_{t}(j) = \sum_{i \ne j} p_t(i) k_{i \to j}
- p_t(j) \left[ \sum_{i \ne j} k_{j \to i} \right]
\label{master}
\end{eqnarray}
In this section, we describe how the results of the previous sections should be adapted for this case.

\subsection{Dynamical entropy for the continuous-time master equation}

Let us first recall how the master Equation (\ref{master}) can be obtained as the continuous limit $\Delta t \to 0$ 
 of the Markov chain of Eq. \ref{markovmicrononeq}.
In the limit where the elementary time step 
becomes small $\Delta t \to 0$, it is natural to assume that the transition
probability $w_{i \to j} $ from $i$ to another state $j \ne i$ becomes 
proportional to $\Delta t$
\begin{eqnarray}
  w_{i \to j } \opsimeq_{ \Delta t \to 0} k_{i \to j} \Delta t
\label{Wwij}
\end{eqnarray}
where $k_{i \to j}$ represents the { \it transition rate per unit-time}  from $i$ to $j$.
The normalization of Eq. \ref{markovnormaw} yields that the transition
probability to remain on site $i$ takes the form
\begin{eqnarray}
   w_{i \to i} =1-\sum_{j \ne i}  w_{i \to j}
\opsimeq_{\Delta t  \to 0} 1-  \Delta t k_{out}(i)
\label{Wwiout}
\end{eqnarray}
where 
\begin{eqnarray}
 k_{out}(i) \equiv  \sum_{j \ne i} k_{i \to j} 
\label{wouti}
\end{eqnarray}
 represents the { \it total exit rate } out of state $i$.
In this limit $ \Delta t \to 0$, the finite-time Markov chain of Eq. \ref{markovmicrononeq}
thus becomes the master Equation \ref{master}.

However as explained in \cite{vivien}, the continuous time description
leads to some difficulties when one considers the dynamical entropy of Eq. \ref{straj}.
  Indeed, the naive continuous limit of Eq. \ref{hKSmarkov} becomes
\begin{eqnarray}
\frac{S_{dyn}(t)}{t} 
&& = - \frac{1}{\Delta t}  \sum_{i} p^{st}(i) \sum_{j} w_{i \to j}  \ln  w_{i \to j}
 \nonumber \\ && 
\opsimeq_{\Delta t \to 0}
 - \frac{1}{\Delta t}  \sum_{i} p^{st}(i) \left[ \sum_{j \ne i}  k_{i \to j} \Delta t
\ln (k_{i \to j} \Delta t) 
+  \left(1- \Delta t k_{out}(i)   \right)
\ln \left(1- \Delta t  k_{out}(i)   \right)  \right]
 \nonumber \\
&& \opsimeq_{\Delta t \to 0}
 -   \sum_{i} p^{st}(i)
\left[ \sum_{j \ne i}  k_{i \to j} 
\ln (k_{i \to j} \Delta t ) 
-  k_{out}(i)    \right]
 \label{hKSmaster}
\end{eqnarray}
i.e. the elementary time $\Delta t$ remains in the argument of the logarithm
to have a non-dimensional quantity.
This difficulty to define the 'absolute' entropy in problems involving continuous variables
can be usually circumvented by considering the 'relative' entropy with respect to some reference.
In our present framework, the reference will be the corresponding equilibrium master equation
defined by transition rates $k^{eq}_{i \to j}$ satisfying the detailed balance condition (Eq \ref{detailed})
\begin{eqnarray}
 p^{eq}(i) k^{eq}_{i \to j} =  p^{eq}(j) k^{eq}_{j \to i}
\label{detailedk}
\end{eqnarray}

The continuous-time limit $\Delta t \to 0$ of the relative dynamical entropy of Eq. \ref{strajrelmarkov} reads
\begin{eqnarray}
 \frac{S^{dyn}_{rel} (t) }{t} && = - \frac{1}{\Delta t} \sum_i p^{st}(i) \sum_j 
 w_{i \to j} \ln \left( \frac{w_{i \to j}}{w^{eq}_{i \to j}} \right) \nonumber \\
 && = - \frac{1}{\Delta t}  \sum_i p^{st}(i) \left[  \sum_{j \ne i}  k_{i \to j} \Delta t  \ln \left( \frac{k_{i \to j}}{k^{eq}_{i \to j}} \right) 
 +  \left( 1-  \Delta t k_{out}(i) \right) \ln \left( \frac{1-  \Delta t k_{out}(i)}{1-  \Delta t k^{eq}_{out}(i) } \right)\right]
  \nonumber \\
 &&  \opsimeq_{\Delta t \to 0}    \sum_i p^{st}(i) \left[  - \sum_{j \ne i}  k_{i \to j}  \ln \left( \frac{k_{i \to j}}{k^{eq}_{i \to j}} \right) 
 +  k_{out}(i) -  k^{eq}_{out}(i)  \right] \nonumber \\
 &&  \opsimeq_{\Delta t \to 0}    \sum_i p^{st}(i) \left[  - \sum_{j \ne i}  k_{i \to j}  \ln \left( \frac{k_{i \to j}}{k^{eq}_{i \to j}} \right) 
 +  \sum_{j \ne i} k_{i \to j}  -  \sum_{j \ne i} k^{eq}_{i \to j}   \right]
\label{strajrelmaster}
\end{eqnarray}
So this relative dynamical entropy is well-defined for continuous time Master Equations, and can be used
in maximization procedures.

\subsection{Maximization of the relative entropy with the appropriate constraints }

We wish to maximize the relative dynamical entropy of Eq. \ref{strajrelmaster} with respect to all stationary states 
$p^{st}(i)$ satisfying the normalization (as in Eq. \ref{normapst})
\begin{eqnarray}
{\cal N} \equiv \sum_i p^{st}(i) =1
\label{normapstmaster}
\end{eqnarray}
and with respect to all possible transition rates 
$k_{i \to j}$ with $j \ne i$ (note that here there is no normalization equation as Eq \ref{nimarkovnorma},
since it has already been taken into account in Eqs \ref{Wwiout} and \ref{wouti}) that have $p^{st}(i)$ as stationary distribution
(equivalent of Eq. \ref{sistmarkov})
\begin{eqnarray}
\Sigma_i \equiv \sum_{j \ne i} p^{st}(j) k_{j \to i} - p^{st}(i) \sum_{j \ne i} k_{i \to j} =0
\label{sistmaster}
\end{eqnarray}
in the presence of the following flux constraint (equivalent of Eq. \ref{javmicro})
\begin{eqnarray}
\frac{J^{dyn}(t)}{t} \equiv \sum_{i} p^{st}(i) \sum_{j \ne i} k_{i \to j} J_{i \to j} =J_0
\label{jenerj0master}
\end{eqnarray} 

So we introduce the following functional with appropriate Lagrange multipliers
\begin{eqnarray}
\Psi && \equiv \frac{S^{dyn}_{rel} (t)}{t} 
- \rho({\cal N} -1) 
- \sum_i \mu_i \left(\Sigma_i  \right)
+ \nu \left( \frac{J^{dyn}(t)}{t} - J_0  \right)
\nonumber \\
&& =  \sum_i p^{st}(i) \left[  - \sum_{j \ne i}  k_{i \to j}  \ln \left( \frac{k_{i \to j}}{k^{eq}_{i \to j}} \right) 
 +  \sum_{j \ne i} k_{i \to j}  -  \sum_{j \ne i} k^{eq}_{i \to j}   \right]
- \rho(\sum_i p^{st}(i) -1) 
\nonumber \\
&& 
- \sum_i 
\mu_i \left(  \sum_{j \ne i} p^{st}(j) k_{j \to i} - p^{st}(i) \sum_{j \ne i} k_{i \to j} \right)
+ \nu \left( \sum_{i} p^{st}(i) \sum_{j \ne i} k_{i \to j} J_{i \to j} - J_0  \right)
\label{psimaster}
\end{eqnarray}

The optimization with respect to $ k_{i \to j}$ with $i \ne j$ 
\begin{eqnarray}
0= \frac{ \delta \Psi }{\delta k_{i \to j}  }
 =  p^{st}(i) \left[  -     \ln \left( \frac{k_{i \to j}}{k^{eq}_{i \to j}} \right)   - \mu_j + \mu_i + \nu J_{i \to j}  \right]
\label{psimasteroptik}
\end{eqnarray}
yields 
\begin{eqnarray}
    k_{i \to j} = k^{eq}_{i \to j}  e^{  \mu_i - \mu_j  + \nu J_{i \to j}  }
\label{psimasteroptiksolu}
\end{eqnarray}
which is the appropriate continuous limit of the Markov chain result of Eq. \ref{solumicro} 
and of the Evans result cited in Eq. \ref{wdrivenevans}.

The optimization with respect to $p^{st}(i)$
\begin{eqnarray}
0= \frac{ \delta \Psi }{\delta p^{st}(i)  }
&& =    \sum_{j \ne i}  k_{i \to j}  \left[ - \ln \left( \frac{k_{i \to j}}{k^{eq}_{i \to j}} \right)  +1 -  \mu_j + \mu_i  + \nu J_{i \to j}  \right]
    -  \sum_{j \ne i} k^{eq}_{i \to j}   - \rho
 \label{psimasteroptip}
\end{eqnarray}
can be simplified using Eq. \ref{psimasteroptik} into
\begin{eqnarray}
0 =    \sum_{j \ne i}  k_{i \to j}     -  \sum_{j \ne i} k^{eq}_{i \to j}   - \rho
 \label{psimasteroptipsolu}
\end{eqnarray}
or equivalently in terms of the total exit rates introduced in Eq \ref{wouti}, one obtains
\begin{eqnarray}
k_{out}(i) =   k_{out}^{eq}(i) + \rho
 \label{invariantkout}
\end{eqnarray}
Physically, this means that for all configurations $i$, the exit rate is shifted by the same amount $\rho$
with respect to the equilibrium case.
Eq. \ref{invariantkout} has been previously obtained by Baule and Evans in \cite{evansinvariants} within their slightly different perspective,
in order to obtain a set of invariant quantities for shear flows and to compute the non-equilibrium rates 
via some network rules (see  \cite{evansinvariants} for more details).

In contrast to Eq. \ref{psimasteroptik}, Eq. \ref{invariantkout} can appear as 'new' with respect to the discrete-time Markov chain 
result of Eq. \ref{solumicro}.
Another apparent difference is that the discrete-time  Markov chain 
result of Eq. \ref{solumicro} contains both the stationary distribution $p^{st}(i)$ and the transition probabilities $w_{i \to j}$
that are determined together, whereas here the two types of equations (Eq. \ref{psimasteroptik} and Eq. \ref{invariantkout})
 concern only the transition rates, and $p^{st}(i)$ does not appear anymore.
 However, this absence of $p^{st}(i)$ is only apparent, because $p^{st}(i)$ is of course directly determined by the rates $k_{i \to j}$
 via the stationarity condition of Eq. \ref{sistmaster}.
 To make clearer the correspondence with the discrete-time Markov chain result of Eq. \ref{solumicro},
 it is thus useful to reformulate the continuous-time master equation solution as follows.
 
 \subsection{Reformulation of the solution in terms of an eigenvalue problem }
 
 The aim of this section is to reformulate the solution found above in Eqs \ref{psimasteroptik} and \ref{invariantkout}
 in terms of an eigenvalue problem to see the similarity with the discrete-time Markov chain solution of Eq.  \ref{solumicro}.
 
 To replace the matrix $M$ of Eq. \ref{solumicroM}, it is convenient to introduce the matrix $N$ defined by the matrix elements
\begin{eqnarray}
N_{i \ne j} && = k^{eq}_{i \to j} e^{\nu J_{i \to j} } \nonumber \\
N_{i i} && = - \sum_{j \ne i} k^{eq}_{i \to j} \equiv - k^{eq}_{out} (i)
\label{solumasterN}
\end{eqnarray}
Setting $e^{-\mu_j}= <j \vert R>$, Eq. \ref{psimasteroptik} becomes
\begin{eqnarray}
    k_{i \to j} = N_{ij} \frac{<j \vert R> }{<i \vert R>}
\label{solumasterk}
\end{eqnarray}
whereas Eq. \ref{invariantkout} becomes
\begin{eqnarray}
 \rho=  k_{out}(i) -   k_{out}^{eq}(i)  = \sum_{j \ne i} N_{ij} \frac{<j \vert R> }{<i \vert R>} + N_{ii} 
 = \frac{ 1 }{<i \vert R>} \sum_j <i \vert N \vert j><j \vert R>= \frac{ <i \vert N \vert R> }{<i \vert R>} 
\label{solumasterkout}
\end{eqnarray}
meaning that $\vert R>$ is a right eigenvector of the matrix $N$ corresponding to the eigenvalue $\rho$
\begin{eqnarray}
    N \vert R> =  \rho \vert R>
\label{solumasterR}
\end{eqnarray}
Let us now rewrite the stationary equation of Eq. \ref{sistmaster} using Eq. \ref{solumasterN}, Eq. \ref{solumasterk}
and Eq. \ref{invariantkout}
\begin{eqnarray}
0=\sum_{i \ne j} p^{st}(i) k_{i \to j}  - p^{st}(j) k_{out} (j)
= \sum_{i \ne j} p^{st}(i)  N_{ij} \frac{<j \vert R> }{<i \vert R>} - p^{st}(j) \left(  k^{eq}_{out}(j) + \rho \right)
\label{sistmasterbis}
\end{eqnarray}
It is thus convenient to set
\begin{eqnarray}
 p^{st}(i)  = <i \vert R> <L \vert i>
\label{pstmasterRL}
\end{eqnarray}
to rewrite Eq. \ref{sistmasterbis} as 
\begin{eqnarray}
0= \sum_{i \ne j}  <L \vert i> N_{ij}  +  <L \vert j> \left( N_{jj}  -  \rho \right)= \sum_{i }  <L \vert i> <i \vert N \vert j > - \rho <L \vert j>
= <L \vert  N \vert j > - \rho <L \vert j>
\end{eqnarray}
meaning that $<L \vert$ is a left eigenvector of the matrix $N$ corresponding to the eigenvalue $\rho$ 
\begin{eqnarray}
  <L \vert N =   \rho <L \vert 
\label{solumasterL}
\end{eqnarray}

In summary, the solution of Eqs \ref{psimasteroptik} and \ref{invariantkout}
for the continuous-time master equation can be written  in terms of an eigenvalue problem for the matrix $N$ 
introduced in Eq. \ref{solumasterN} as follows : 
\begin{eqnarray}
 p^{st}(i)  && = <i \vert R> <L \vert i> \nonumber \\
 k_{i \to j} && = N_{ij} \frac{<j \vert R> }{<i \vert R>}
\label{solumastermicro}
\end{eqnarray}
where $\vert R>$ and $<L \vert$ are the right and the left positive eigenvectors associated with the largest
eigenvalue $\rho$ of the matrix $N$, with the following normalization
\begin{eqnarray}
 1 =  <L \vert R> 
\label{solumastermicronorma}
\end{eqnarray}
The correspondence with the discrete-time Markov chain result of Eq. \ref{solumicro} is now clear.
As a final remark, let us mention that in the equilibrium case $\nu=0$, the matrix $N $ is the generator of the equilibrium dynamics,
 so the maximal eigenvalue corresponds to $  \rho_{eq} =0$,
the corresponding right and the left eigenvectors being simply $<i \vert R_{eq}>=1$ and  $ <L_{eq} \vert i >=p^{eq}(i) $.

\subsection{ Relation with the 'constrained dynamics' introduced to study large deviations }

The reformulation in terms of an eigenvalue problem described in the previous section
is also useful to make the link with the the 'constrained dynamics' that have been
introduced in various studies concerning large deviations of stochastic processes
described by master equations
(see for instance the recent works \cite{damien1,jack,popkov,damien2} and references therein). In this context, the point of view is as follows : one considers some 'true' dynamics and one is interested into large deviations properties of some observable like the current. To compute the probability of rare events giving rise to an anomalous current for the 'true' dynamics, it is useful to introduce an 'auxiliary' dynamics that takes into account the conditioning to produce a given anomalous current
(see \cite{damien1,jack,popkov,damien2} for more details).
The transition rules of this auxiliary dynamics and its stationary state 
are given in terms of left/right eigenvectors of a modified operator
via the same formula written above, see for instance 
Eqs (2.15) (2.17) (2.18) (2.19) in \cite{damien1} or Eqs (2.16) (2.18) and (2.19) in \cite{jack} : the exchange of the roles of the right and left eigenvectors with respect to the present notations comes from the different choice in the writing of matrices, since \cite{damien1,jack} have chosen the 'quantum mechanical' convention where the initial state is on the right and the final state is on the left.
 The fact that exactly the same equations appear can be understood
as follows. In the present paper, we have described how the maximization of the dynamical entropy yields Eq. \ref{soluoptipsitrajrel} at a formal level, 
and give the results of the previous section
when applied to an explicit master equation. 
But of course another possibility could be to take the formal solution of Eq. \ref{soluoptipsitrajrel} as a starting point, and to derive the consequences of this reweighting of trajectories for the specific case of a master equation : this is exactly 
the route followed in the large deviation studies mentioned above \cite{damien1,jack,popkov,damien2} and so one should indeed obtain the same results by consistency.
However, even if the equations are exactly the same, the physical interpretation
of this modified dynamics is different : in the large deviation studies  \cite{damien1,jack,popkov,damien2}, this modified dynamics is usually presented only as a useful technical tool to better understand the large deviations of the 'true' dynamics,
whereas in the approach summarized in the present paper, one considers that 
the modified dynamics is the 'real' dynamics in the presence of an imposed current.

An interesting output of this comparison with large deviation studies
is that it provides specific studies of this modified dynamics
in models different from the examples considered in the works of Evans
  \cite{evans,evansinvariants,evansetal,evansreview} :
the auxiliary dynamics corresponding to the large deviations of the current
of the Asymmetric Simple Exclusion Process has been
 studied in various situations or regimes 
in \cite{damien1,popkov,damien2}, whereas constrained dynamics for the Glauber Ising chain has been considered in \cite{jack}. Since the change of measure is a standard tool in the large deviation theory, many other examples can actually be found in the huge number of works concerning large deviations for stochastic processes.

\section{Conclusions and Perspectives }

\label{sec_conclusion}

In this paper, we have argued that if one wishes to formulate
a general principle based on the maximization of some notion of 'entropy' for
non-equilibrium steady states, the Shannon entropy associated
to the probability distribution of dynamical trajectories
is definitely the most natural, as first proposed by Filyokov and Karpov in 1967.
The general idea is then to maximize dynamical entropy of Eq. \ref{straj}
in the presence of appropriate constraints, including the macroscopic current of interest, via the method of Lagrange multipliers.
We have tried to give a self-contained and unified presentation of this type of approach.
We have first described at the formal level how this maximization leads to  generalized Gibbs distribution for the probability 
distribution of dynamical trajectories, and to some fluctuation relation of the integrated current.
We have then discussed in detail the cases of well defined stochastic dynamics generated either by discrete-time Markov Chains
or by continuous-time Master Equations. In the cases where the use of the 'full' dynamical entropy of Eq. \ref{straj} leads to some difficulties,
we have shown how the use of the 'relative' dynamical entropy of Eq. \ref{strajrel} allows to solve the problems.
The obtained results are in full agreement with the Evans approach 
called 'Non-equilibrium Counterpart to detailed balance' \cite{evans,evansinvariants,evansetal,evansreview},
but give a slightly different perspective. In particular, we have obtained that the stationary distribution and the transitions probabilities
or transitions rates could be obtained in various cases from an eigenvalue problem. 
Finally, we have explained the link with the constrained dynamics 
that are often introduced in large deviations studies of stochastic processes.

In this article, we have remained at a very general level with 
a dynamics visiting a series of configurations, to see more clearly what
general properties could emerge, since it is at this level of generality that
the statistical physics theory of equilibrium is formulated.
 However it is clear that it would be interesting
in the future to study more precisely 
how this type of approach can be applied in various models of interest,
and to discuss whether the obtained dynamics can be considered as 'real'.
An important issue is of course to compare
with other approaches that solve exactly some non-equilibrium
dynamics. In the field of quantum models, the non-equilibrium steady states
have been found to be generalized Gibbs states, but they involve an infinite number
of conserved quantities coming from the integrability of the
considered models \cite{ogata,dragi}. For non-integrable models, one may thus hope that the non-equilibrium steady states correspond to much simpler generalized Gibbs states.

As a final remark, we should stress that the 
idea of maximizing some dynamical entropy to describe non-equilibrium steady states
has been argued here to hold for 'physical systems' for which, in the absence of any 
imposed current, the notion of equilibrium exists and corresponds to the 
maximization of the usual 'static' entropy of statistical physics,
and for which effective stochastic models satisfying detailed balance
are well accepted to describe the equilibrium dynamics.
But of course, in the field of non-equilibrium, many models of interest
are inspired not by physics, but by 
biology (like predator-prey models),   
sociology (like road traffic models), politics (like voter models) etc...
In all these cases, it is clear that one can choose arbitrarily the microscopic
rates to model the considered phenomenon as one wishes.

\section*{Acknowledgements}

It is a pleasure to thank 
the Saclay working group 'Climate and Statistical mechanics', where
  B. Dubrulle, C. Herbert and D. Paillard have given talks about their work
concerning the 'maximum entropy production' \cite{herbert} :
this led me to read the review \cite{revueprodentropie} where I have found the papers of Filyokov and Karpov  \cite{russes,russeseq,russesneareq}.
I also wish to thank C. Appert-Rolland, T. Bodineau , R. Ch\'etrite, Mike Evans, R. Jack, W. Hoover, V. Lecomte, C. Maes, K. Mallick, D. Simon, P. Sollich, F. van Wijland and B. Wynants for useful correspondence, discussion or for indicating to me relevant references.

\appendix

 \section{ Reminder on the maximization of the static entropy for the equilibrium state } 

\label{static}

At equilibrium, one is interested into the equilibrium probabilities 
$P^{eq}(i)$ of occupations of configurations $i$. The Shannon entropy associated with the equilibrium distribution is
\begin{eqnarray}
S^{eq} = -\sum_i P^{eq}(i) \ln P^{eq}(i)
\label{seq}
\end{eqnarray}

(i) If the only constraint is the normalization,
\begin{eqnarray}
N \equiv \sum_i P^{eq}(i) =1
\label{normapeq}
\end{eqnarray}
the maximization of the entropy of Eq. \ref{seq} with the constraint
of Eq. \ref{normapeq} can be obtained by introducing a Lagrange multiplier
$\lambda$ and the functional
\begin{eqnarray}
\Phi_{\lambda}  = S^{eq} - \lambda(N-1) 
= -\sum_i P^{eq}(i) \ln P^{eq}(i)- \lambda(\sum_i P^{eq}(i) -1) 
\label{phieqN}
\end{eqnarray}
The optimization with respect to $P_{eq}(j)$ yields
\begin{eqnarray}
0= \frac{ \delta \Phi_{\lambda} }{\delta P_{eq}(j)}  
= - \ln P^{eq}(j)- 1- \lambda
\label{optiN}
\end{eqnarray}
leading to $P_{eq}(j)=e^{-1-\lambda}$, where $\lambda$ is determined by the normalization constraint of Eq. \ref{normapeq}. 
So $ P_{eq}(j)$ is simply uniform over all states.

(ii) If in addition to the normalization of Eq. \ref{normapeq},
one imposes also a constraint for the averaged energy
\begin{eqnarray}
<E> \equiv \sum_i E_i P^{eq}(i) = E_0
\label{eaveq}
\end{eqnarray}
one has to introduce another Lagrange multiplier $\beta$
and the functional
\begin{eqnarray}
\Phi_{\lambda,\beta}  && = S^{eq} - \lambda(N-1) - \beta(<E>-E_0 )
\nonumber \\
&& = -\sum_i P^{eq}(i) \ln P^{eq}(i)- \lambda(\sum_i P^{eq}(i) -1) 
- \beta(\sum_i E_i P^{eq}(i)-E_0 )
\label{phieqNE}
\end{eqnarray}
The optimization with respect to $P_{eq}(j)$ yields
\begin{eqnarray}
0= \frac{ \delta \Phi_{\lambda,\beta} }{\delta P_{eq}(j)}  
= - \ln P^{eq}(j)- 1- \lambda - \beta E_j
\label{optiNE}
\end{eqnarray}
leading to $P_{eq}(j)=e^{-1-\lambda- \beta E_j}$, 
where $\lambda$ is determined by the normalization constraint
 of Eq. \ref{normapeq}, and $\beta$ by Eq. \ref{eaveq}
Then $ P_{eq}(j)$ follows the Boltzmann-Gibbs distribution
\begin{eqnarray}
P_{eq}(j) = \frac{ e^{- \beta E_j}}{ \sum_i e^{- \beta E_i}}
\label{Boltzmann-Gibbs}
\end{eqnarray}

This derivation of the Boltzmann-Gibbs distribution, 
introduced by Jaynes \cite{jaynes57}, has the advantage 
to be very simple and to really show what the principle of maximum entropy
contains. For more details on the interest to consider statistical physics
from the point of view of the Shannon information entropy,
we refer the reader to the 'old' book \cite{brillouin} 
and to recent presentations \cite{balian} and references therein.

 \section{ Technical simplifications for an alternate Markov chain  } 

\label{sec_alter}

In their original paper \cite{russes}, Filyokov and Karpov have not considered the homogeneous Markov chain of Eq. \ref{markov}, but have focused instead onto an alternate Markov chain,
that has been also  reconsidered recently by Favretti \cite{favretti}.
In this Appendix, we show how this alternate Markov chain framework
 yields technical simplifications with respect
to the solution described in section \ref{sec_noneq} concerning the homogeneous Markov chain.

\subsection{ Alternate Markov Chain } 

In this section, we consider the alternate Markov chain, where the transition probabilities
$W_{i \to j}$ takes alternatively two sets of values $A_{i \to j}$ and $B_{i \to j}$, 
so that the probability of a trajectory
of Eq. \ref{pomegamarkov} now becomes
\begin{eqnarray}
{\cal P}(\Omega_{[0,2t]}=\{i_0,i_1,...,i_{2t} \})
= P^{even}(i_0) A_{i_0 \to i_1}  B_{i_1 \to i_2} ... 
A_{i_{2t-2} \to i_{2t-1}} B_{i_{2t-1} \to i_{2t}}
\label{pomegamarkovab}
\end{eqnarray}
with the normalizations (as in Eq. \ref{markovnormaw})
\begin{eqnarray}
{\cal N}^{A}_i && \equiv  \sum_j  A_{i \to j} =1 \nonumber \\
{\cal N}^{B}_i && \equiv  \sum_j  B_{i \to j} =1
\label{markovnormaab}
\end{eqnarray}
The stationary states $P^{even}(i)$, $P^{odd}(i)$ at even and odd times
associated with this alternate Markov chain
satisfy
\begin{eqnarray}
\Sigma_i^{even}  \equiv   \sum_j P^{odd}(j) B_{j \to i} - P^{even}(i) =0 \nonumber \\
\Sigma_i^{odd}  \equiv  \sum_j P^{even}(j) A_{j \to i}- P^{odd}(i) =0
\label{stmarkovab}
\end{eqnarray}
with the normalizations 
\begin{eqnarray}
{\cal N}^{even} \equiv \sum_i P^{even}(i) && =1 \nonumber \\
{\cal N}^{odd} \equiv \sum_i P^{odd}(i) && =1
\label{normastmarkovab}
\end{eqnarray}
The dynamical entropy of Eq. \ref{strajmarkovfinal} becomes
\begin{eqnarray}
 S^{dyn}(2t)  
= - t  \left[ \sum_{i} P^{even}(i) \sum_{j} A_{i \to j} \ln  A_{i \to j}
+ \sum_{j} P^{odd}(i) \sum_{j} B_{i \to j} \ln  B_{i \to j}
\right]
\label{strajmarkovfinalAB}
\end{eqnarray}
The averaged energy is now fixed by the constraint
\begin{eqnarray}
\frac{E^{dyn}(2t)}{2t} \equiv \frac{1}{2} \left[\sum_{i} P^{even}(i) E_{i} + \sum_{i} P^{odd}(i) E_{i}\right] = E_0
\label{eavab}
\end{eqnarray}
that replaces Eq. \ref{eav}.

\subsection{ Simplification concerning the current constraint }

In  \cite{russes,favretti}, the interest was in an energy current flowing
through the system : 
the advantage of this formulation with an alternate Markov chain
is that one may consider that during the evolution described by $A$,
there is a flow of energy entering the system, 
whereas during the evolution described by $B$,
there is a flow of energy coming out of the system.
To remain in a stationary state, these two flows have to be equal,
and can be simply expressed in terms of the difference of averaged energies
between the even and odd stationary states
\begin{eqnarray}
J_E^{in} &&  = \sum_i P^{even}(i) \sum_{j } A_{i \to j} (E(j)-E(i)) = \sum_j E(j) P^{odd}(j) - \sum_i E(i) P^{even}(i) \nonumber \\
J_E^{out} &&  = \sum_i P^{odd}(i) \sum_{j } B_{i \to j} (E(i)-E(j)) =  \sum_i E(i) P^{odd}(i)
- \sum_j E(j) P^{even}(j) = J_E^{in}
\label{jenergyab}
\end{eqnarray}
In the following, we consider more generally the case of a current $J_K$
 that can be written similarly
as the difference between some observable $K$ between
the even and odd stationary states
($K$ is the 'charge' associated with the current $J_K$)
\begin{eqnarray}
J_K^{in} &&  = \sum_i P^{even}(i) \sum_{j } A_{i \to j} (K(j)-K(i)) = \sum_j K(j) P^{odd}(j) - \sum_i K(i) P^{even}(i) \nonumber \\
J_K^{out} &&  = \sum_i P^{odd}(i) \sum_{j } B_{i \to j} (K(i)-K(j)) =  \sum_i K(i) P^{odd}(i)
- \sum_j K(j) P^{even}(j) = J_K^{in}
\label{jenerab}
\end{eqnarray}
so that the constraint of Eq. \ref{jenerj0}  can be replaced by\begin{eqnarray}
\frac{J^{dyn}(2t)}{t} \equiv  \sum_i K(i) \left[ P^{odd}(i) -  P^{even}(i) \right]
\label{jenerj0ab}
\end{eqnarray}

\subsection{ Optimization of the Lagrange functional }

So the functional of Eq. \ref{psimarkovfull} is replaced by
(some factors of $2$ have been added to simplify slightly the notations)
\begin{eqnarray}
&& \Psi  \equiv \frac{S^{dyn}(2t)}{t} 
- \rho^{even}({\cal N}^{even} -1) - \rho^{odd}({\cal N}^{odd} -1) 
- \beta(\frac{E^{dyn}(t)}{t}-E_0 )
\nonumber \\
&& - \sum_i \lambda_i^A \left({\cal N}^A_i -1 \right)- \sum_i \lambda_i^B \left({\cal N}^B_i -1 \right)
- \sum_i \mu_i^{even} \left(\Sigma_i^{even}  \right)- \sum_i \mu_i^{odd} \left(\Sigma_i^{odd}  \right)
+ \nu \left(  \frac{J^{dyn}(2t)}{t}- J_0  \right)
\nonumber \\
&& =   - \left[ \sum_{i} P^{even}(i) \sum_{j} A_{i \to j} \ln  A_{i \to j}
+ \sum_{j} P^{odd}(i) \sum_{j} B_{i \to j} \ln  B_{i \to j}
\right]
\nonumber \\
&&
- \rho^{even}(\sum_i P^{even}(i)
 -1) - \rho^{odd}(\sum_i P^{odd}(i) -1) 
 \nonumber \\ &&
- \beta \left(  \left[\sum_{i} P^{even}(i) E_{i} + \sum_{i} P^{odd}(i) E_{i}\right] -2 E_0 \right)
- \sum_i \lambda_i^A \left(\sum_j  A_{i \to j} -1 \right)- \sum_i \lambda_i^B \left(\sum_j  B_{i \to j} -1 \right) \nonumber \\
&&
- \sum_i \mu_i^{even} \left(\sum_j P^{odd}(j) B_{j \to i} - P^{even}(i)
  \right)- \sum_i \mu_i^{odd} \left(\sum_j P^{even}(j) A_{j \to i}- P^{odd}(i)
  \right)\nonumber \\
&&
+ \nu \left( \sum_i K(i) \left[ P^{odd}(i) -  P^{even}(i) \right] - J_0  \right)
\label{psimarkovfullab}
\end{eqnarray}

The optimization with respect to $A_{i \to j}$ 
\begin{eqnarray}
0= \frac{ \delta \Psi }{\delta A_{i \to j}  }  =   P^{even}(i) 
 \left[  - \ln  A_{i \to j} -1 - \lambda_i^A -  \mu_j^{odd}  \right]
\label{psimarkovfullaboptia}
\end{eqnarray}
yields 
\begin{eqnarray}
  A_{i \to j} = e^{-1 - \lambda_i^A -  \mu_j^{odd} }
\label{solua}
\end{eqnarray}
The normalization of Eq. \ref{markovnormaab} yields that $\lambda_i^A$ is independent of $i$ and given by
\begin{eqnarray}
e^{1 + \lambda_i^A  }  = \sum_j e^{ -  \mu_j^{odd} }
\label{solulamdaa}
\end{eqnarray}
The stationary Equation of Eq. \ref{stmarkovab}   yields using Eq. \ref{normastmarkovab}
\begin{eqnarray}
P^{odd}(j)=  \sum_i P^{even}(i) A_{i \to j} =  (\sum_i P^{even}(i)) \frac{e^{ -  \mu_j^{odd} }}{ \sum_k e^{ -  \mu_k^{odd} }} = \frac{e^{ -  \mu_j^{odd} }}{ \sum_k e^{ -  \mu_k^{odd} }}
\label{stmarkovabuse}
\end{eqnarray}
The optimization with respect to  $B_{i \to j}$ yields similar equations and solutions,
so that finally one has the simple forms
\begin{eqnarray}
  A_{i \to j} = P^{odd}(j) = \frac{e^{ -  \mu_j^{odd} }}{ \sum_k e^{ -  \mu_k^{odd} }} \nonumber \\
 B_{i \to j} = P^{even}(j) = \frac{e^{ -  \mu_j^{even} }}{ \sum_k e^{ -  \mu_k^{even} }}
\label{soluab}
\end{eqnarray}

The optimization of Eq. \ref{psimarkovfullab} with respect to $P^{even}(i)$ yields 
\begin{eqnarray}
0 && = \frac{ \delta \Psi }{\delta P^{even}(i)  } =   -  \sum_{j} A_{i \to j} \ln  A_{i \to j}
- \rho^{even}
- \beta  E_{i} + \mu_i^{even}  - \sum_j \mu_j^{odd}  A_{i \to j} - \nu  K(i) 
\label{psimarkovfullaboptipeven}
\end{eqnarray}
yields using Eq \ref{soluab}
\begin{eqnarray}
- \mu_i^{even} = - \rho^{even} - \beta  E_{i}  - \nu  K(i)  +\ln \left( \sum_k e^{ -  \mu_k^{odd} } \right)
\label{mieven}
\end{eqnarray}
Eq. \ref{soluab} then yields
\begin{eqnarray}
 P^{even}(i) = \frac{e^{ -  \mu_i^{even} }}{ \sum_k e^{ -  \mu_k^{even} }}
= \frac{e^{  - \beta E_i - \nu K_i   }}{ \sum_k e^{ - \beta E_k - \nu K_k  }}
\label{solupeven}
\end{eqnarray}
Similarly, the optimization with respect to $P^{even}(i)$
yields 
\begin{eqnarray}
 P^{odd}(i) = \frac{e^{ - \beta E_i + \nu K_i    }}{ \sum_k e^{  - \beta E_k +\nu K_k  }}
\label{solupodd}
\end{eqnarray}

So here in contrast to the case of the homogeneous Markov chain 
discussed in section \ref{sec_noneq}, one obtains two Boltzmann-Gibbs distributions,
without any prefactor like the functions  $(y_i,z_i)$ in Eq \ref{pstwnoneq}.
This simple result was found on the special case of the energy flow $K_i=E_i$
by Favretti \cite{favretti} ( who has used the correct expression for the Kolmogorov-Sinai entropy
of the alternate Markov chain, whereas an erroneous expression has been actually used in the 
initial work \cite{russes} ) : the even and odd stationary distributions of Eq. 
\ref{solupeven} and \ref{solupodd} are then two Boltzmann-Gibbs distributions
at two different temperatures that are fixed by the constraints on the averaged energy
and on the averaged energy flow.

Despite the technical simplifications of the alternate Markov chain,
we find that it is rather artificial from a physical point of view.
Since the result means that the system oscillates between two distinct Boltzmann-Gibbs distributions,
we feel that it could only correspond to physical situations
where the energy is added or removed instantaneously at even and odd times,
and that the system then relaxes to its new equilibrium during the
macroscopic time $\tau$ of the Markov chain.
Otherwise, if the energy were added and removed continuously during the time intervals,
this would mean that one should somehow have thermal equilibrium at all times
with an adiabatic change of temperature, which seems extremely restrictive.
This is why in the present paper, we have chosen to put the main focus of the
case of the homogeneous Markov Chain.

\end{document}